\documentclass[aps,prd,nofootinbib,reprint,superscriptaddress]{revtex4-1}

\usepackage{blindtext}
\usepackage{amsfonts}
\usepackage{tensor}
\usepackage{graphicx}
\usepackage{pict2e}
\usepackage{balance}

\usepackage{amssymb, amsmath,environ}

\usepackage{color}


\newcommand{\bse}{\begin{subequations}}
	\newcommand{\ese}{\end{subequations}}
\newcommand{\be}{\begin{equation}}
\newcommand{\ee}{\end{equation}}

\NewEnviron{eqsplit}{%
	\begin{equation}\begin{split}
	\BODY
	\end{split}\end{equation}
}

\makeatletter
\newcommand*\bigcdot{\mathpalette\bigcdot@{.5}}
\newcommand*\bigcdot@[2]{\mathbin{\vcenter{\hbox{\scalebox{#2}{$\m@th#1\bullet$}}}}}
\makeatother

\newcommand{\bea}{\begin{eqnarray}}
\newcommand{\eea}{\end{eqnarray}}
\newcommand{\ba}{\begin{array}}
	\newcommand{\ea}{\end{array}}

\newcommand{\K}{\mathcal{K}}

\newcommand{\A}{\mathcal{A}}

\newcommand{\B}{\mathcal{B}}

\newcommand{\D}{\mathcal{D}}

\newcommand{\la}{\langle}
\newcommand{\ra}{\rangle}

\begin{document}

	\title{Systematic Analysis of Flow Distributions}

	%
	%
	%
	%
	%
	%
	%
	%
	
	\author{Hadi Mehrabpour}
	\email[]{hadi.mehrabpour.hm@gmail.com}
	\affiliation{School of Particles and Accelerators, Institute for Research in Fundamental Sciences (IPM), P.O. Box 19395-5531, Tehran, Iran}
	\affiliation{Frankfurt Institute for Advanced Studies, Giersch Science Center, \\D-60438 Frankfurt am Main,Germany}
	
\begin{abstract}
The information of the event-by-event fluctuations are extracted from flow harmonic distributions and cumulants, which can be done experimentally. In this work, we employ the standard method of Gram-Charlier series with normal kernel to find such distribution, which is the generalization of recently introduced flow distributions for the studies of the event-by-event fluctuations. In this path, we find the shifted cumulants $j_n\{2k\}$ which are consist of the collision geometry information. 
The experimental data imply that not only all of the information about the event-by-event fluctuations of collision zone properties and different stages of heavy ion process are not encoded in the radial flow distribution $p(v_n)$, but also the observables describing harmonic flows can generally be given by the joint distribution $\mathcal{P}(v_1,v_2,...)$. In such way, we first introduce a set of joint cumulants $\K_{nm}$, and then we find the flow joint distribution using these joint cumulants. Finally, we show that the Symmetric Cumulants $SC(2,3)$ and $SC(2,4)$ obtained from ALICE data are explained by the combinations $\K_{22}+\frac{1}{2}\K_{04}-\K_{31}$ and $\K_{22}+4\K_{11}^2$.

\end{abstract}
	\maketitle
\section{Introduction}\label{sec1} 
The collective behavior of the initial fireball, which is created in heavy-ion collisions, can be experimentally measured by anisotropic flow. Anisotropic flow is traditionally quantified with harmonics $v_n$,
the coefficients of the momentum distribution Fourier expansion in the azimuthal direction, which have been measured by the several experimental groups at Relativistic Heavy Ion Collider (RHIC) and Large Hadron Collider (LHC) \cite{Ackermann:2000tr,Lacey:2001va,Park:2001gm,Aamodt:2010pa,ALICE:2011ab,Chatrchyan:2012ta,ATLAS:2011ah,Aad:2014vba}. Due to the randomness of reaction plane angle and low statistic at each event, anisotropic flow finding is experimentally challenging. There are several techniques to solve these problems \cite{Poskanzer:1998yz,Borghini:2001vi,Borghini:2000sa,Bhalerao:2003yq,Bhalerao:2003xf}. One of them is 2k-particle correlation functions $c_n\{2k\}$ (radial cumulants) \cite{Borghini:2000sa,Borghini:2001vi}.
On the other hand, the experimental results show that the flow harmonics fluctuate event-by-event even if a specific centrality class is considered \cite{Adler:2007aa,Aamodt:2010cz}. The flow fluctuations contain the information of the collision geometry, quantum fluctuations at initial state, and effects of different evolution stages in heavy-ion process \cite{Schenke:2012wb,Miller:2003kd}. The distribution of flow harmonic not only can solve the problems of the reaction plane angle effect and low statistic in a given event, but also it can help us extract the information of observed event-by-event fluctuations. So, these issues motivate us to study the radial flow distributions $p(v_n)$. 

Experimentally flow distributions for second, third, and forth harmonics  have been obtained using the unfolding method \cite{Jia:2013tja,Aad:2013xma}. Also it has been found that the Bessel Gaussian distribution describe the observed flow distributions in some centrality collisions \cite{Aad:2013xma,Voloshin:2006gz,Voloshin:2007pc}.

It should be noted that the information of flow fluctuations not only are encoded in the flow harmonic distribution $p(v_n)$, but also these information can be extracted from radial cumulants $c_n\{2k\}$ \cite{navid,hadi}. Consequently, finding a right set of cumulants and connecting them to flow harmonic distribution $p(v_n)$ can help us get closer to an exact interpretation of the event-by-event fluctuations.
Thereby, different distributions and their cumulants have recently been introduced and investigated to explain the contributions of all evolution stages on the fluctuations. In Ref. \cite{navid}, odd flow harmonic distributions have been obtained by employing two dimensional standardized cumulants. In addition, using Gram-Charlier A series with orthogonal polynomials, $p_{odd}(v_n)$ has been found in Ref. \cite{hadi}.  
The experimental data of the even flow harmonics can not be explained by the Bessel Gaussian distribution in peripheral collisions. So, finding the corrections to the Bessel Gaussian distribution is crucial. Ref. \cite{hadi} considered an ansatz series as the corrections to the Bessel Gaussian distribution . They employed moments to find the corresponding coefficients of this series. Their suggested flow distribution could decently explain both even and odd harmonics.       

Experiments show that the event-plane correlations and event-by-event correlations of flow magnitudes are non-vanishing \cite{ALICE:2016kpq,Jia:2012sa,Aad:2014fla}. Thus, all of the information about  the fluctuations can be extracted from a joint flow distribution $\mathcal{P}(v_1,v_2,...)$, which can explain the correlations between flow harmonics, event-by-event initial fluctuations, and correlations between different stages in heavy-ion collision processes.    
Now a question arises: Is there an unambiguous technique to find such radial flow distributions $p(v_n)$? Furthermore, can we find a joint flow distribution to interpret the most general form of the event-by-event fluctuations? The purpose of this paper is the answer to this question by introducing a systematic analysis of flow fluctuations so that we can find the cumulant coefficients and the consequently flow harmonic distributions. 

In this work, we employ the standard method of Gram-Charlier series with normal kernel to introduce this analysis in Sec. \ref{sec2}. Also, we show that using this technique, one can find the radial cumulants $c_n\{2k\}$ to flow moments $\la v_n^{2k}\ra$.   
In Sec. \ref{sec3}, expanding the relation between moment and cumulant characteristic functions to 2-dimension, we first rederive the relations between $\la v_n^{2k}\ra$ and $c_n\{2k\}$, and then we find the distribution of odd flow harmonics which has been found in Refs. \cite{navid} and \cite{hadi}. 
After that, we find a general form of flow distribution which is true for both even and odd harmonics. In the way to find this distribution, we firstly find the shifted cumulants $j_n\{2k\}$ which are consist of the collision geometry information, and then using the standard method of finding Gram-Charlier series, we obtain the flow distribution.
In the final step, we introduce a joint distribution of flow harmonics and its cumulants $\K_{nm}$ in Sec. \ref{sec4}. We conclude Sec. \ref{sec4} by showing ALICE data can be described the combinations of joint cumulants. Moreover, the simulation data can be explained by the obtained joint distribution of flow harmonics. We present the conclusion in Sec. \ref{con}.
\section{systematic technique}\label{sec2}
Azimuthal asymmetry of the final state single-particle distribution, 
\begin{equation}\label{eq1}
\frac{dN}{d\phi}= \frac{1}{2\pi}\sum_{n=-\infty}^{\infty}V_n e^{-in\phi},
\end{equation}
is quantified by the complex anisotropic flow coefficients (or flow vector) $V_n\equiv v_n e^{in\psi_n}=\{e^{in\phi}\}$, where $\phi$ is the azimuthal direction of an emitted particle, $v_n$ is the amplitude of anisotropic flow in the $n$th harmonic, and $\psi_n$ is the corresponding symmetry plane.
Anisotropic flow, which is the hydrodynamic response to the anisotropic initial density profile, is one of the most important observables in characterizing the properties of QGP evolution. 
Flow fluctuates event by event, because it is stochastic and fluctuations are unavoidable. It is worth mentioning that the flow event-by-event fluctuations are a reflection of the initial state fluctuations such that it is sensitive to details of initial geometry and its fluctuations. All of these lead us to search for the underlying $\text{p.d.f.}$ of flow fluctuations.
Hence, presenting a general method to find the flow distribution and its cumulants to explain the event-by-event flow fluctuations becomes important. In this section, we introduce such method using the relation between moment and cumulant generating function. 
For simplicity we consider one dimensional generating functions. In statistics, the generating function of moments in one dimensional is $G(t)=\int dx\;e^{itx}p(x)\equiv\la e^{itx}\ra$. Also, the cumulant-generating function is defined as the logarithm of the characteristic function, $K(t)\equiv \ln\la e^{itx}\ra =\sum_{n=1}^{\infty}\frac{(it)^n}{n!}\kappa_n$, which implies
$G(t)=\exp[{\sum_{n=1}^{\infty}\frac{(it)^n}{n!}\kappa_n}]$\cite{Kendall:1945,Cramer:1999,Krzanowski:2000}. Note that $\kappa_n$ is the $n$th cumulants.
Furthermore, the relation between cumulants and moments by using definitions of $G(t)$ and $K(t)$ is
\begin{equation}\label{eq4} 
1+\sum_{n=1}^{\infty}\frac{\mu_n (it)^n}{n!}=\exp\Big(\sum_{n=1}^{\infty}\frac{\kappa_n (it)^n}{n!} \Big),
\end{equation}
where $\mu_n=\la x^n\ra$.
The relation between $n$th moment and cumulants can be obtained by differentiating both sides of Eq.(\ref{eq4}) $n$ times and evaluating the result at $t=0$,
\begin{equation}
K^{(n)}(t)|_{t=0}=(\log G(t))^{(n)}|_{t=0}.
\end{equation}
Let us expand $G(t)$ in Eq.(\ref{eq4}) to second order and set $\kappa_1=\mu_1\equiv\mu$ and $\kappa_2=\sigma^2$. The structure of the generating function, thus, becomes the following
\begin{equation}\label{eq5} 
\begin{aligned}
G(t)&=\exp[{\sum_{n=3}^{\infty}\kappa_n\frac{(it)^n}{n!}}+\kappa_1(it)+\kappa_2\frac{(it)^2}{2!}]
\\&=\exp[\sum_{n=3}^{\infty}\kappa_n\frac{(it)^n}{n!}]e^{it\mu-\frac{t^2\sigma^2}{2}},
\\&\equiv\exp[\sum_{n=3}^{\infty}\kappa_n\frac{(it)^n}{n!}]G_N(t),
\end{aligned}
\end{equation}
with $G_N(t)\equiv \exp\left[it\mu-\frac{t^2\sigma^2}{2}\right]$.
Note that integrating by parts gives $(it)^n G_N(t)$ as the characteristic function of $(-\boldsymbol{D})^n G_N(x)$, where $\boldsymbol{D}$ is the differential operator \footnote{Consider a function $p(x)$ with Fourier transform $G(t)$ such that
	$$p(x)=\frac{1}{2\pi}\int_{-\infty}^{\infty}G(t)e^{-itx}dt.$$ To find the Fourier transform of $dp(x)/dx$, a simple way, using the anti-transform:
	$$\frac{dp(x)}{dx}=\frac{d}{dx}\Big(\frac{1}{2\pi}\int_{-\infty}^{\infty}G(t)e^{-itx}dt\Big)=\frac{1}{2\pi}\int_{-\infty}^{\infty}(-it)G(t)e^{-itx}dt.$$Hence, the Fourier transform $dp(x)/dx$ is $(-it)G(t)$.
}. On the other hand, we can find the probability density function $p(x)$ by using the last line of defined moment-generating function in Eq.(\ref{eq5}):
\begin{equation}\label{qq3} 
\begin{aligned}
p(x)&=\frac{1}{2\pi}\int dt\;e^{-itx}G(t)
\\&\approx\frac{e^{-\frac{(x-\mu)^2}{2\sigma^2}}}{\sqrt{2\pi}\sigma}\Big(1+\sum_{n=3}^{\infty}\frac{\kappa_n}{n!\sigma^n}He_{n}(\frac{x-\mu}{\sigma})\Big),
\end{aligned}
\end{equation}
where $He_n$ is the probabilists' Hermite polynomials,
\begin{equation}
He_n(x)=(-1)^ne^{\frac{x^2}{2}}\frac{d^n}{dx^n}e^{-\frac{x^2}{2}}.
\end{equation} 
This technique is the standard method of finding Gram-Charlier series with the normal kernel \cite{Brenn:2017}. In this method, one can find the probability density function (p.d.f) without any considered ansatz for the p.d.f.

To see how this method can help us to find the distribution of flow harmonics and cumulants, we first define the form of characteristic function using Eq.(\ref{eq4}) as follows  
\begin{equation}
G(\boldsymbol{\lambda})=\la e^{i \boldsymbol{v}\cdot\boldsymbol{\lambda}}\ra=\la e^{iv_n\lambda\cos(\Psi_n-\Psi_{\lambda})}\ra, 
\end{equation}
where we have used the notation $\Psi_n=n\psi_n$. Since one-dimensional characteristic function is needed to find the relations between cumulants and moments in the case of flow harmonics, we can integrate over $\Psi_{n}$ to have $G(\lambda)$ \cite{hadi},
\begin{equation}
G(\lambda)=\la J_0(\lambda v_n)\ra.
\end{equation}
So, the relation between the generating functions of 2k-particle cumulants $c_n\{2k\}$ \cite{Borghini:2000sa,Borghini:2001vi} and flow magnitude moments $\la v_n^{2k}\ra$ are
\begin{equation}\label{q2}
\begin{aligned} 
\la J_0(\lambda v_n)\ra&=1+\left(\sum _{k=1}^{\infty } \frac{(-1)^k\lambda ^{2 k} \left\langle v_n^{2 k}\right\rangle }{4^k(k!)^2}\right)\\
&=\exp\Big(\sum _{k=1}^{\infty } \frac{i^{2k}c_n\{2 k\} \lambda ^{2 k}}{4^k(k!)^2}\Big),
\end{aligned}
\end{equation}
where $J_{\nu}$ is the Bessel functions of the first kind,  
\begin{equation}
J_\nu(x)=\sum_{k=0}^{\infty} \frac{(-1)^k}{k!\Gamma(k+\nu+1)}(\frac{x}{2})^{2k+\nu}.
\end{equation}
In the results, 2k-particle cumulants $c_{n}\{2k\}$ can be given to the measured $v_{n}$ at each event by differentiating both sides of Eq.(\ref{q2}) at $\lambda=0$:
\begin{equation}\label{eq6} 
\begin{aligned}
c_n\{2\}&=\left\langle v_n^2\right\rangle ,\\
c_n\{4\}&=\left\langle v_n^4\right\rangle -2 \left\langle v_n^2\right\rangle^2 ,\\
c_n\{6\}&=12 \left\langle v_n^2\right\rangle {}^3-9 \left\langle v_n^4\right\rangle  \left\langle v_n^2\right\rangle +\left\langle v_n^6\right\rangle , \\ 
c_n\{8\}&=-144 \left\langle v_n^2\right\rangle {}^4+144 \left\langle v_n^4\right\rangle  \left\langle v_n^2\right\rangle {}^2\\&\quad\;-16 \left\langle v_n^6\right\rangle  \left\langle v_n^2\right\rangle -18 \left\langle v_n^4\right\rangle {}^2+\left\langle v_n^8\right\rangle , \\
\vdots
\end{aligned}
\end{equation}
As can be seen in Eq.(\ref{eq6}), the odd moment of radial flow distribution are absent in the definitions of $c_n\{2k\}$. So, we can conclude that the radial flow distribution has more information than 2k-particle correlation functions. Finding the flow distribution is left to next section.
 
So far we have presented a well-known technique in statistic theory to find the probability distribution and its cumulants. As can be seen, using this technique we could find the 2k-particle cumulants. In the following we first obtain the flow distribution of odd harmonics \cite{navid}. Then we try to find a general probability distribution to explain the event-by-event fluctuation which is true for both odd and even flow harmonics. 
\section{two dimensional cumulant and moment relations}\label{sec3}
As mentioned earlier, introducing a method to find flow harmonic distribution that extract the maximum amount of information is necessary. Here, we present a technique commonly used in statistics to achieve our goal. To find the relations between moments and cumulants of flow harmonics, we use the joint generating functions \cite{young:2009}, 
\begin{equation}\label{q3}
\log \langle e^{\lambda z + \lambda^* z^*}\rangle =\sum_{k,l}\frac{\lambda^{*k}\lambda^l}{k!l!}\kappa\{k,l\},
\end{equation}
where $\kappa\{k,l\}$ are joint cumulants.
It is worth emphasizing that Eq.(\ref{q3}) is a general formula. Moreover, to find the desired flow distributions we need to modify Eq.(\ref{q3}) by choosing different definitions of $z$ and $\lambda$.

In Ref. \cite{navid}, an expansion of flow distribution for odd harmonics has been found (also see Eq.(24) in Ref. \cite{hadi}). To reproduce this expansion, we have to set $z\equiv V_n$ and $\lambda\equiv(\lambda_x - i\lambda_y)/2$ in Eq.(\ref{q3}). By replacing these considerations in Eq.(\ref{q3}), we have   
\begin{equation}\label{q4}
\begin{aligned}
\la& e^{v_{n,x}\lambda_x + v_{n,y}\lambda_y}\ra
\\&=\exp\left[\sum_{kl}\frac{(\lambda_x+i\lambda_y)^{k}(\lambda_x-i\lambda_y)^l}{2^{(k+l)}k!l!}c_n\{k,l\}\right].
\end{aligned}
\end{equation}
Here we use the common notation of $c_n$ for 2k-particle cumulants. Note that in Eq.(\ref{q4}) only terms with $k=l$ are non-zero. Also, setting $k=l$ the relations in Eq.(\ref{eq6}) are reproduced
\footnote{Note that to find the averaged flow magnitude we have to integrate the generating moments over $\phi_{\lambda}$ in polar coordinates $$G(\boldsymbol{\lambda})=\frac{1}{2\pi}\int_{0}^{2\pi}d\phi_{\lambda} <e^{\boldsymbol{\lambda}.\boldsymbol{v}}>,$$ where $\lambda_x=\lambda\cos\phi_{\lambda}$ and $\lambda_y=\lambda\sin\phi_{\lambda}$.}.
The cumulant $c_n\{k,k\}\equiv c_n\{2k\}$ can be obtained by differentiating both sides of Eq.(\ref{q3}),  
\begin{equation}
\begin{aligned}
	\frac{\partial^{2k}}{\partial\lambda_x^k\partial\lambda_y^k}\Big(\la& e^{v_{n,x}\lambda_x + v_{n,y}\lambda_y}\ra
	\\&=\exp\left[\sum_{kl}\frac{(\lambda_x+i\lambda_y)^{k}(\lambda_x-i\lambda_y)^k}{4^{(k)}(k!)^2}c_n\{2k\}\right]\Big),
\end{aligned}
\end{equation}
and evaluating the results at $\lambda_x=0$ and $\lambda_y=0$. 
To find the odd flow distributions\footnote{Note that $\la v_{n,x}\ra$ is zero for the odd flow distribution.}, we use the first line of Eq.(\ref{qq3}) \footnote{In the integration we use $\lambda_x \to -i \lambda_x$ and $\lambda_y \to -i \lambda_y$.} and the Fourier transformation of characteristic function, $\lambda_x^2 +\lambda_y^2 \to \partial_x^2+\partial_y^2$. The probability distribution for odd harmonics, thus, becomes (see Appendix \ref{app0})     
\begin{equation}
\begin{aligned}
p_{odd}&(v_{n,x},v_{n,y}) \\&=\exp\left[\sum_{k=2}\frac{c_n\{2k\}(\partial_x^2+\partial_y^2)^{k}}{2^{2k}(k!)^2}\right]\left[\frac{1}{\pi c_n\{2\}}e^{\frac{-v_{n,x}^2-v_{n,y}^2}{ c_n\{2\}}}\right].
\end{aligned}
\end{equation}
If we rewrite this distribution in polar coordinates, $v_n^2=v_{n,x}^2+v_{n,y}^2$, we can obtain the radial odd flow distribution,
\begin{equation} 
\begin{aligned}
\int dv_{n,x}&dv_{n,y}\;p_{odd}(v_{n,x},v_{n,y})
\\&=\int\frac{v_ndv_nd\Psi_n}{\pi c_n\{2\}}\exp\left[\sum_{k=2}\frac{c_n\{2k\}\boldsymbol{D}_{v_n,\Psi_n}^{k}}{4^{k}(k!)^2}\right]e^{\frac{-v_n^2}{c_n\{2\}}}\\
&=\int dv_n\; p_{odd}(v_n),
\end{aligned}
\end{equation} 
where $\boldsymbol{D}_{v,\Psi}$ represent $\boldsymbol{D}_v+(1/v^2)\partial_{\Psi}^2$ and $\boldsymbol{D}_{v}$ is $\partial_v^2+(1/v)\partial_v$. Therefore, the radial distribution of odd flow harmonics $p_{odd}(v_n)$ is 
\begin{equation}\label{q7}
p_{odd}(v_n)=\frac{2v_n}{c_n\{2\}}\exp\left[\sum_{k=2}\frac{c_n\{2k\}\boldsymbol{D}_{v_n,\Psi_n}^{k}}{4^{k}(k!)^2}\right]e^{\frac{-v_n^2}{c_n\{2\}}}.
\end{equation}
If the first exponential in Eq.(\ref{q7}) is expanded and truncated to the first order, we have 
\begin{equation}
p'_{odd}(v_n)=\frac{2v_n}{c_n\{2\}}[1+\sum_{k=2} \frac{ c_n \{2k\}}{4^k (k!)^2}\boldsymbol{D}_{v_n}^k]e^{-\frac{v_n^2}{c_n\{2\}}}.
\end{equation}
The form of $p'_{odd}(v_n)$ can be found in terms of cumulants by evaluating the $k$th derivative of $\exp(-\frac{v_n^2}{c_n\{2\}})$ (see Appendix \ref{app1}) and letting $2\sigma^2=c_n\{2\}$ for odd harmonics \cite{navid,hadi},
\begin{equation}\label{q23}
p'_{odd}(v_n) = (\frac{v_n}{\sigma^2})e^{-\frac{v_n^2}{2\sigma^2}}\Big[1+\sum_{k=2} \frac{(-1)^k \Gamma^{odd}_{2k-2}}{k!} \mathit{L}_{k} (v_n^2/(2\sigma^2))\Big],
\end{equation}
where $\Gamma^{odd}_{2k-2}=c_n\{2k\}/c_n\{2\}^k$ and $L_k$ is  the Laguerre polynomials,
\begin{equation} 
L_k(x)=\sum_{n=0}^{k}\bigl(\begin{smallmatrix}
k \\ n
\end{smallmatrix} \bigr)\frac{(-x)^n}{n!}.
\end{equation} 
The expansion (\ref{q23}) is exactly the flow distribution found in Ref. \cite{navid} which can explain any event-by-event flow fluctuations of odd harmonics. 

Because $p_{odd}(v_{n,x},v_{n,y})$ is rotationally symmetric ($\bar{v}_{2n+1}\equiv\la v_{2n+1,x}\ra=0$) and consequently the main features of 2D and radial odd flow distribution are the same, obtaining distribution (\ref{q23}) is simple. But this case is not true for even flow harmonics, since $\bar{v}_{2n}\neq0$. This is because even flow distributions are not rotationally symmetric, and reshuffling $(v_{n,x},v_{n,y})$ leads to a partial loss of information of $p_{even}(v_{n,x},v_{n,y})$. 
Hence, the main challenge is to find a radial flow distribution which can give a good approximation of flow fluctuations for even $n$ so that the least amount of information is lost. 

In the following, we begin to find the flow harmonic distribution and its cumulants by assuming non-zero $\bar{v}_n$.  
Modifying the relation (\ref{q3}) for even flow harmonics, the relation of moment and cumulant generating functions in 2D with $k=l$ can be rewritten as
\begin{equation}\label{q9}
\la e^{(v_{n,x}-\bar{v}_n)\lambda_x + v_{n,y}\lambda_y}\ra=\exp\left[\sum_{k}\frac{(\lambda_x^2+\lambda_y^2)^{k}}{2^{2k}(k!)^2}j_n\{2k\}\right].
\end{equation}
where we consider $z\equiv V_n-\bar{v}_n$ and $\lambda\equiv\frac{\lambda_x - i\lambda_y}{2}$. We simply use the notation $W_n\equiv V_n-\bar{v}_n$ as a shifted flow vector so that $\la W_n\ra=0$ \footnote{As we know, the averaged  ellipticity  $\tilde{v}_2\equiv\la v_{2,x}\ra$ is a manifestation of the geometrical initial ellipticity for events in a given centrality class irrespective of the fluctuations. In general, we are able to define averaged flow harmonic $\tilde{v}_n\equiv\la v_{n,x}\ra$. Also, we know the average of $v_{n,y}$ is zero,$\la v_{n,y}\ra=0$, because the distribution of $v_{n,y}$ is centered at $0$ due to parity  conservation and symmetry with respect to the reaction plane \cite{Giacalone:2016eyu}. Moreover, we defined $z$ with the shifted flow vector $W_n=(v_{v,x}-\bar{v}_n)+iv_{n,y}$.}. The reason for choosing $k=l$ is to avoid obtaining complex $j_n$ cumulants. By differentiating both sides of Eq.(\ref{q9}) at $\lambda_x=0$ and $\lambda_y=0$, one can find the relations between $j_n\{2k\}$ and moments,       
\begin{equation}\label{eq12}
	\begin{aligned}
	j_n\{2\}&=\la w_n^2\ra,\\
	j_n\{4\}&=\la w_n^4\ra-2\la w_n^2\ra^2,\\
	j_n\{6\}&=\la w_n^6\ra+12\la w_n^2\ra^3-9\la w_n^2\ra\la w_n^4\ra,\\
	j_n\{8\}&=\la w_n^8\ra-144\la w_n^2\ra^4+144\la w_n^4\ra\la w_n^2\ra^2\\&\quad\;-16\la w_n^6 \ra\la w_n^2\ra-18\la w_n^4\ra^2,
	\\\vdots 
	\end{aligned}
\end{equation}
where $w_n^2=|W_n|^2=(v_{n,x}-\bar{v}_n)^2+v_{n,y}^2$. 
As we know, the cumulants are invariant under shifting a random variable. For instance, the cumulant of 2-particle azimuthal correlation can be written as $\la e^{in(\phi_1-\phi_2)}\ra-\la e^{in\phi_1}\ra\la e^{-in\phi_2}\ra$ for the case of non-perfect detector \cite{Borghini:2001vi}. If we shift $\phi_1$ and $\phi_2$ by the same quantity $\phi_i\to\phi_i-\theta$, the cumulant would stay invariant. In Eq.(\ref{eq12}), $j_n\{2k\}$ are consist of some moments which are not invariant under shifting $\phi_i\to\phi_i-\theta$. Of course, one can find the cumulants $j_n\{2k\}$ are shift invariant by removing such moments. However, we renamed $j_n\{2k\}$ as "shifted cumulants" to avoid confusion.
\begin{figure}[t!]
	\begin{center}
		\begin{tabular}{c}
			\includegraphics[scale=0.42]{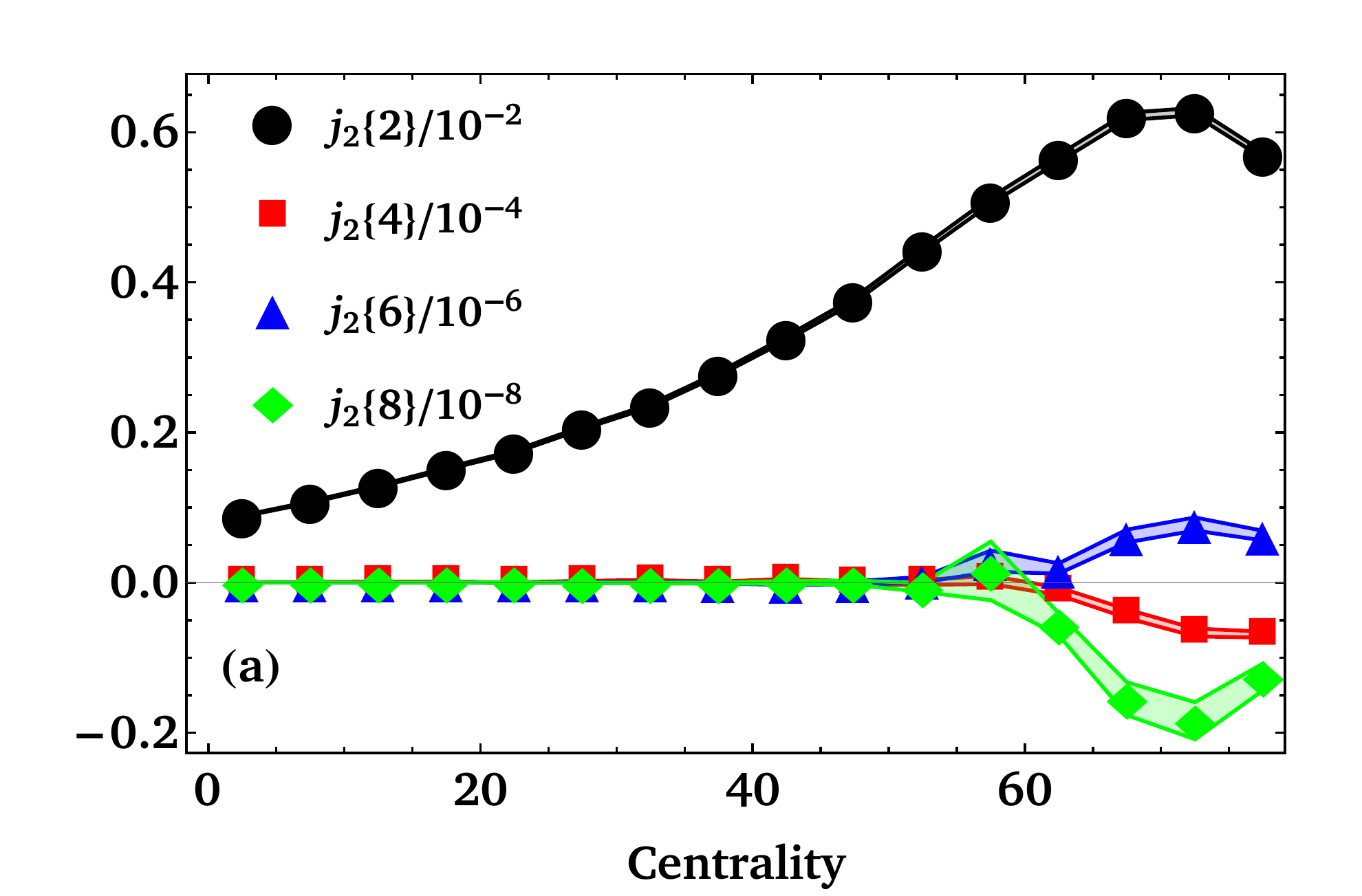}\\			\includegraphics[scale=0.49]{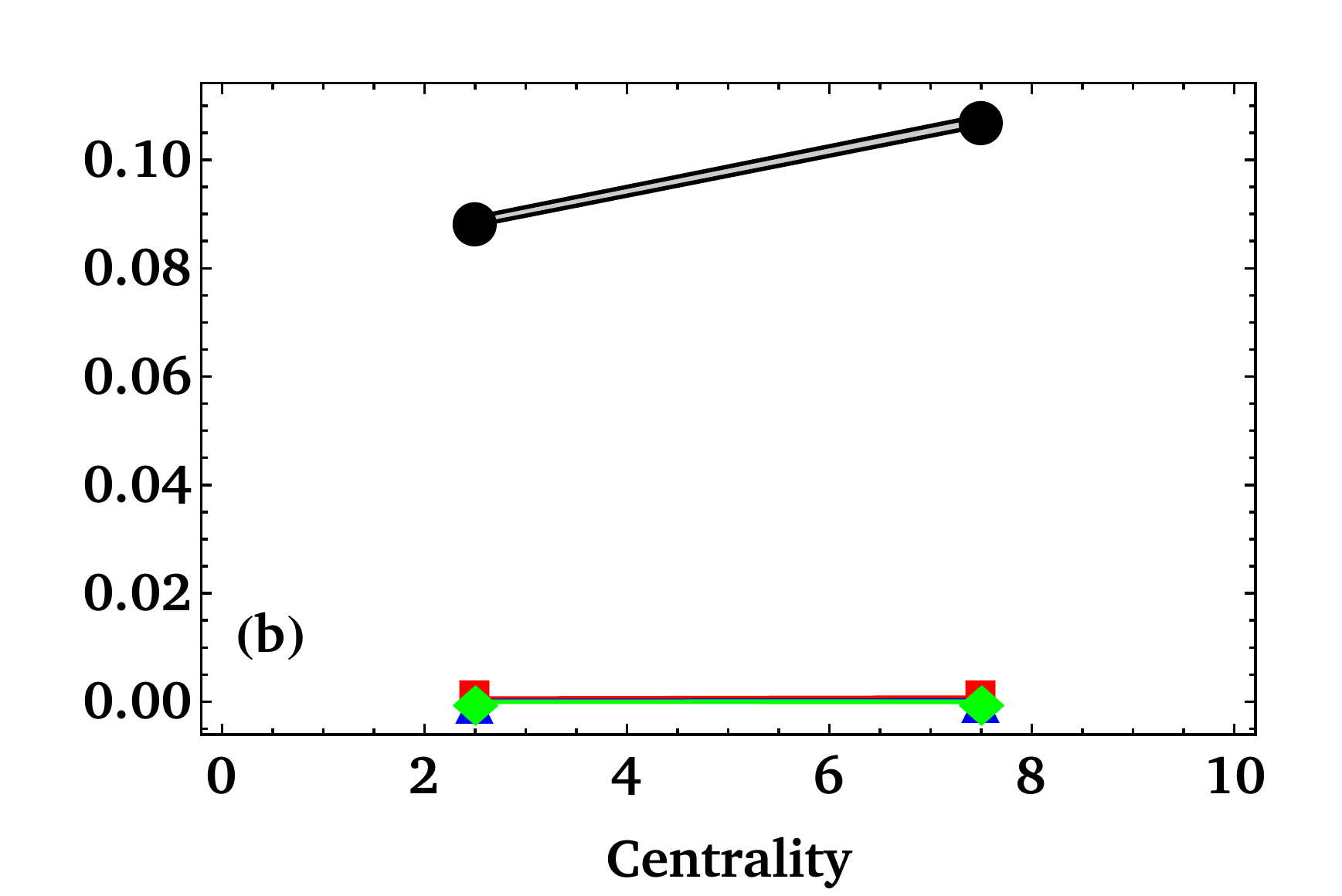}
		\end{tabular}		
		\caption{(Color online) Comparing the cumulants $j_2\{2k\}$ for $k=2$, $3$, and $4$ obtained from the iEBE-VISHNU event generator.
		We separated results in two panels to compare the results in peripheral and most central collisions ($0-5\%$ and $5-10\%$ centrality classes).} 
		\label{Check1}
	\end{center}
\end{figure}
Also, it should be noticed that by choosing $\bar{v}_n=0$, 2k-particle correlation functions $c_n\{2k\}$ can be recovered and we have $j_n\{2k\}=c_n\{2k\}$.

Fig. \ref{Check1} presents the cumulants $j_n\{2k\}$ for $n=2$, obtained from the iEBE-VISHNU output, and the comparisons between them.
It should be mentioned that in the present work, we study Pb-Pb collision with center-of-mass energy per nucleon pair $\sqrt{s_{NN}}= 2.76$ TeV. In this work, we use MC-Glauber model for the initial state such that the wounded nucleon/binary collision mixing ratio was set to $0.118$. The hydrodynamic starting time $\tau_0$ was set to $0.6\;fm/c$, and $\eta/s=0.08$ and zero balk viscosity are used for the hydrodynamic evolution. Note that the events generate in sixteen centrality classes between $0-80\%$ and we generated $14000$ events for each centrality. It should be emphasized that we have taken into account the charged hadrons $\pi^{\pm}$, $K^{\pm}$, $p$ and $\bar{p}$ in the final particle distribution which are in the transverse momentum range $0.28<p_T<4$ GeV.
In Fig. \ref{Check1}, we scaled shifted cumulants as $j_n\{2k\}/10^{-2k}$ to show $j_n\{4\}$, $j_n\{4\}$, $j_n\{6\}$, and $j_n\{8\}$ in a plot. Also, we drew the shifted cumulants in $0-5\%$ and $5-10\%$ centrality classes to compare the results of peripheral and most central collisions.
As demonstrated in this figure, the differences between $j_n\{4\}$, $j_n\{6\}$, and $j_n\{8\}$ are sensible in peripheral central collisions such that the relation between $j_n\{2k\}$ is 
\begin{equation}
j_n\{2\} \gg j_n\{4\}\gg j_n\{6\}\gg j_n\{8\}\gg\cdots.
\end{equation}
We expected this relation, because the experimental results \cite{ATLAS:2011ah} show that $p_{even}(v_n)$ have a deviation from Bessel-Gaussian. This deviation is more pronounced in peripheral collisions where the Bessel-Gaussian distribution can not explain experimental data. Furthermore, we expect that the cumulants $j_n\{2k\}$ can quantify the main features of a distribution near Bessel-Gaussian.
\begin{figure}[t!]
	\begin{center}
		\begin{tabular}{c}
			\includegraphics[scale=0.5]{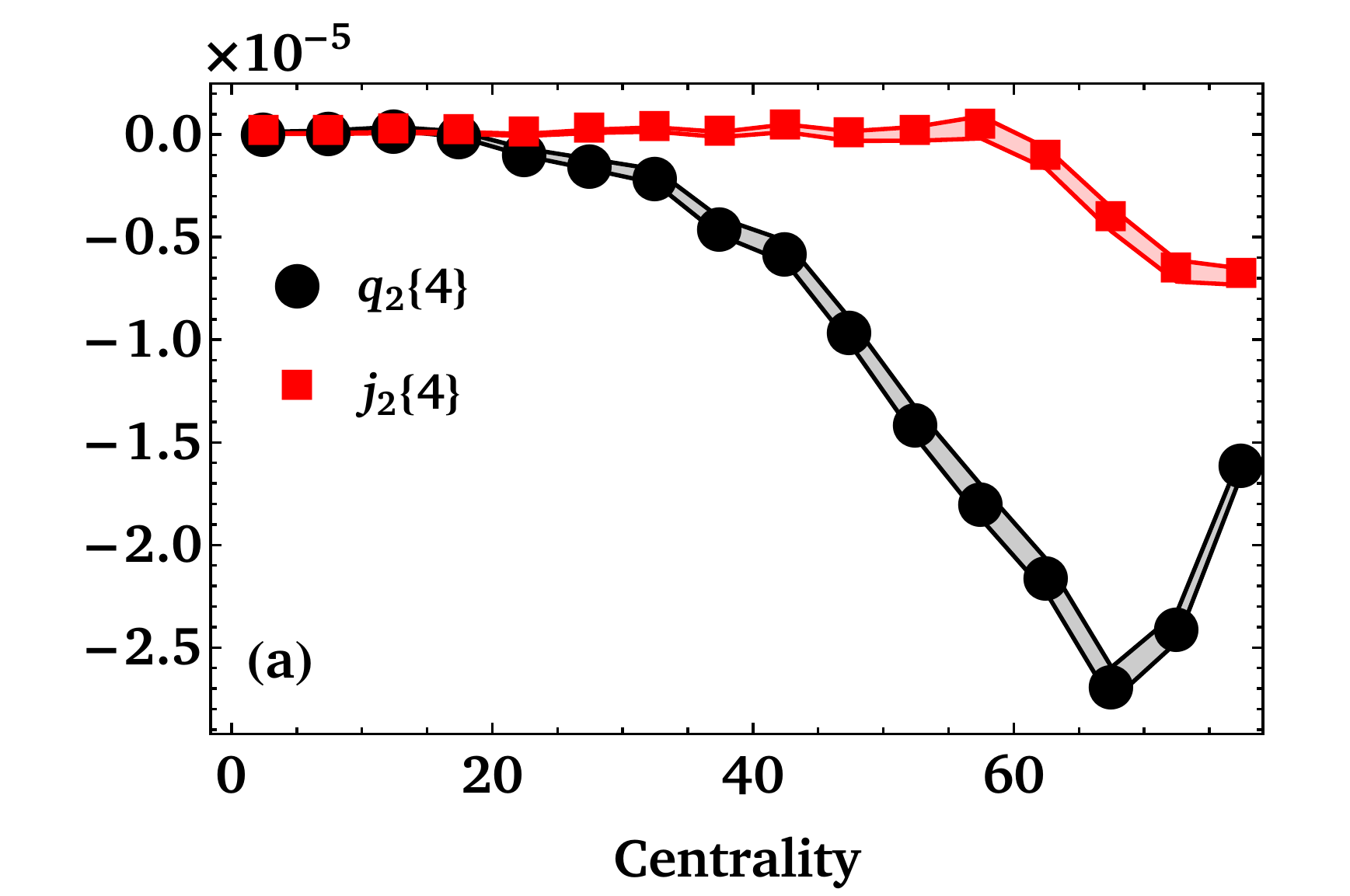}\\			\includegraphics[scale=0.5]{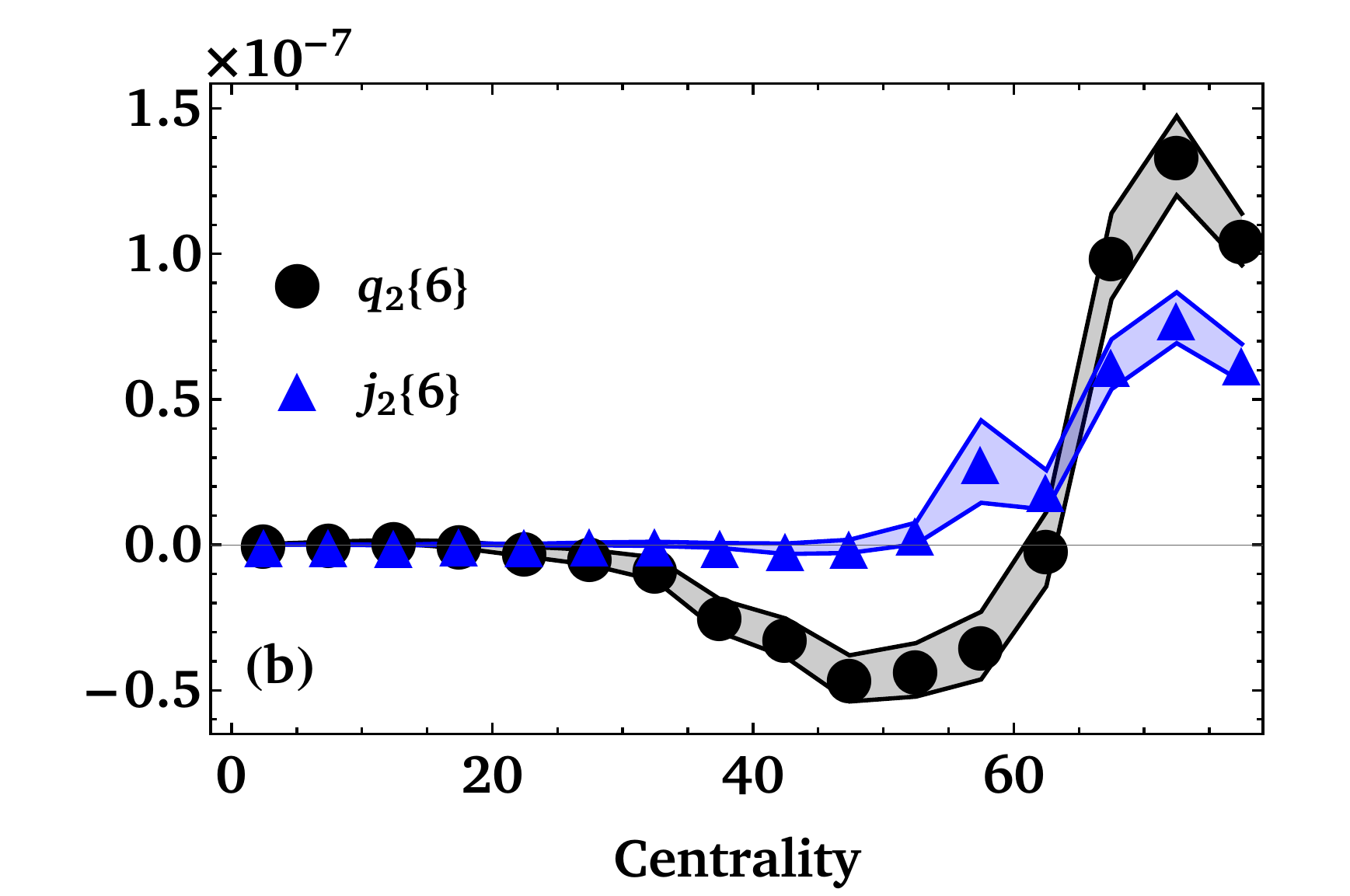}\\ 
			\includegraphics[scale=0.498]{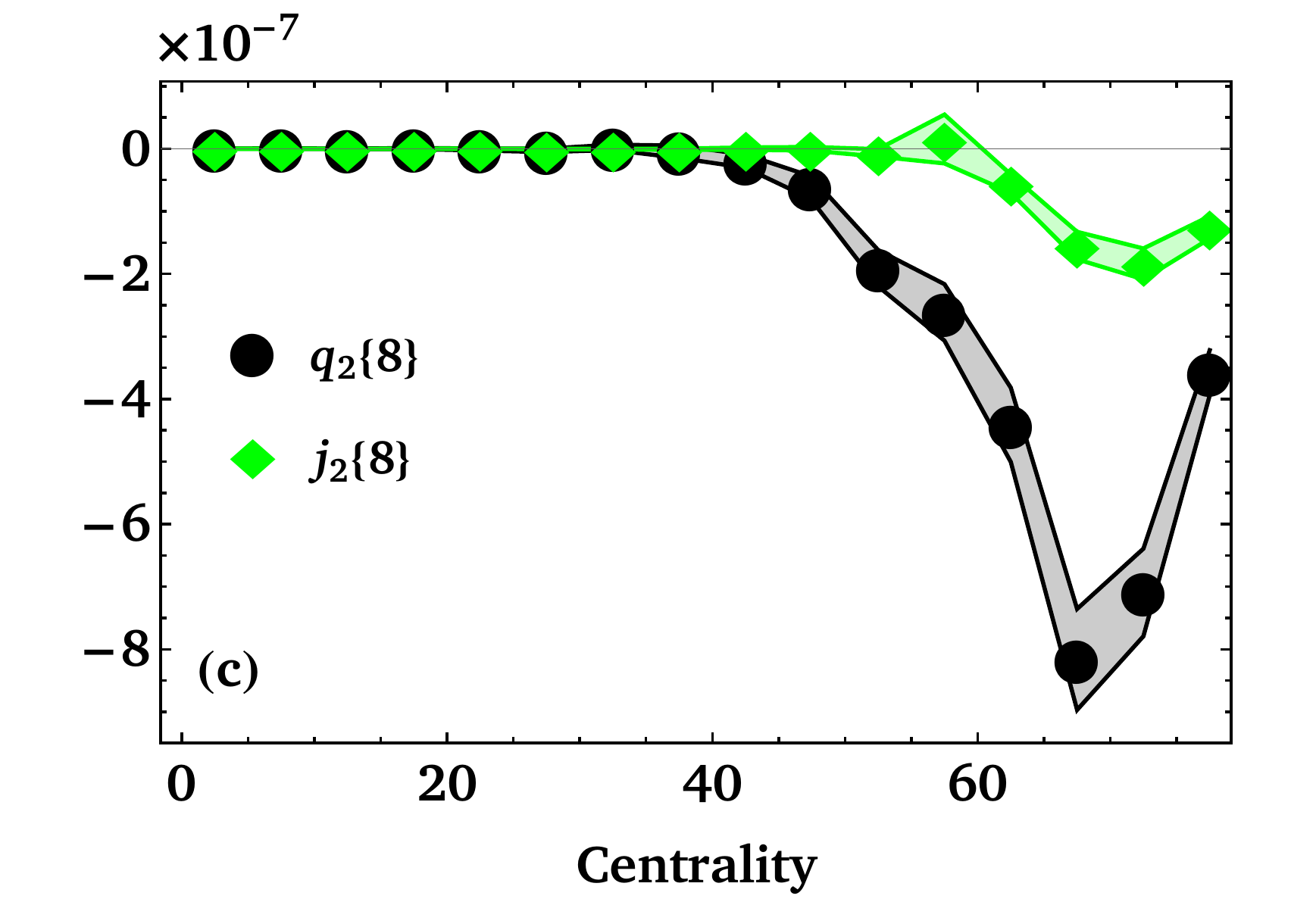}			
		\end{tabular}		
		\caption{(Color online) Comparing the amount of information of cumulants $j_2\{2k\}$ and $q_2\{2k\}$ introduced in Ref. \cite{hadi} as function of centrality.} 
		\label{Check2}
	\end{center}
\end{figure}
In Ref. \cite{hadi}, a new set of cumulants $q_n\{2k\}$ has been defined to study the distributions near Bessel-Gaussian. These cumulants have also been obtained from 2k-particle correlation functions $c_n\{2k\}$. Replacing the definitions of $q_n\{2k\}$ (see Eq.(36) in Ref.\cite{hadi}) in Eq.(\ref{eq12}), one can find  
\begin{equation}\label{eq13}
\begin{aligned}
j_n\{2\}&=q_n\{2\},
\\j_n\{4\}&=\frac{2}{7}\Big(4q_n\{4\}+q_n\{2\}^2+40q_n\{2\}\bar{v}^2+16\bar{v}^4
\\&-4\bar{v}(4\la v_{n,x}^3\ra+5\bar{v}\la v_{n,y}^2\ra+2\la v_{n,x}v_{n,y}^2\ra)
\\&-4\la v_{n,x}^2v_{n,y}^2\ra\Big),
\\\vdots
\end{aligned}
\end{equation}
such that $j_n\{2k\}=q_n\{2k\}+...,\; \text{for}\; k\geq 2$.
A comparison between $j_n\{2k\}$ and $q_n\{2k\}$ obtained from iEBE-VISHNU are presented in Fig. \ref{Check2}. As can be seen, the difference between these sets of cumulants for $k\geq2$ is significant, especially for mid-centralities and peripheral collisions. This means that the amount of encoded information in these two sets are different.  
It should be noted the cumulant set $q_n\{2k\}$ has been defined by using the moments of the radial flow distribution $p_q(v_n;\bar{v}_n)$ in Ref. \cite{hadi}, but here we only used the relation between the joint cumulant and moment generating functions to find $j_n\{2k\}$. This means that our technique does not require any knowledge about the flow distributions.

The main challenge is finding the form of flow harmonic distributions by considering $\bar{v}_n\neq0$. If we obtain the Fourier transformation of joint characteristic function of moments in Eq.(\ref{q9}), $\lambda_x \to -i \partial_x$ and $\lambda_y \to -i \partial_y$ , the 2D distribution $p(v_{n,x},v_{n,y})$\footnote{A flow distribution that works for even harmonics is a general distribution that is true for all harmonics. Furthermore, we use the notation $p(v_{n,x},v_{n,y})$ instead of $p_{even}(v_{n,x},v_{n,y})$ for non-rotational symmetric flow distribution.} is obtained as
\begin{equation}\label{q10}
p(v_{n,x},v_{n,y})=\exp\left[\sum_{k=2}\frac{j_n\{2k\}\boldsymbol{D}^{k}}{4^{k}(k!)^2}\right]\mathcal{F}(v_{n,x},v_{n,y}),
\end{equation}
where $\boldsymbol{D}$ is the differential operator with respect to $\lambda_x$ and $\lambda_y$. Also, the distribution $\sqrt{2\pi}\sigma\mathcal{F}(v_{n,x},v_{n,y})$ is a 2D Gaussian distribution with mean $\bar{v}_n$ and standard deviation $\sqrt{j_n\{2\}/2}$. 
After some calculations in Cartesian coordinates, we have
\begin{equation}\label{qq16} 
\boldsymbol{D}^k\mathcal{F}(v_{n,x},v_{n,y})= \frac{(-1)^k 4^k k!}{j_n\{2\}^k}\mathcal{F}(v_{n,x},v_{n,y})\mathit{L}_k (\frac{w_n^2}{j_n\{2\}}).
\end{equation} 
Since we follow the radial flow distribution, Eq.(\ref{qq16}) can be written in polar coordinates as follows
\begin{equation}\label{qq17}
\begin{aligned} 
\boldsymbol{D}_{v_n,\Psi_n}^k \mathcal{F}(v_n;\bar{v}_n,\Psi_n)&= \frac{(-1)^k 4^k k!}{j_n\{2\}^k} \mathcal{F}(v_n;\bar{v}_n,\Psi_n)
\\&\times\Big(\mathit{L}_k (\frac{v_n^2+\bar{v}_n^2}{j_n\{2\}})+A_k+B_k\Big).
\end{aligned} 
\end{equation}
where the terms of $A_k$ and $B_k$ are
\begin{equation}
A_k=\alpha_k,\quad B_k=\sum_{l=1}^{k}\beta_{kl}\cos l\Psi_n.
\end{equation}
The derivation of the $n$th derivative of $\mathcal{F}(v_{n,x},v_{n,y})$ in Eq.(\ref{qq17}) and definitions of the coefficients $\alpha$ and $\beta$ are in the Appendix \ref{app2}. 
If we integrate ``$\mathcal{F}(v_n;\bar{v}_n,\Psi_n)\cos l\Psi_n$" over $\Psi_n$, we find that
\begin{equation}\label{qq18} 
\begin{aligned} 
\int_{0}^{\infty}&v_ndv_n\int_{0}^{2\pi}d\Psi\;\mathcal{F}(v_n;\bar{v}_n,\Psi_n)\cos l\Psi_n
\\&=\int_{0}^{\infty}dv_n\;(\frac{2v_n}{j_n\{2\}})e^{-\frac{v_n^2+\bar{v}_n^2}{j_n\{2\}}}I_l(\frac{2v_n\bar{v}_n}{j_n\{2\}})\\&=\int_{0}^{\infty}dr\;\mathcal{F}(v_n;\bar{v}_n)I_l(\frac{2v_n\bar{v}_n}{j_n\{2\}}).
\end{aligned}
\end{equation}
Using Eq.(\ref{qq17}), we find that
\begin{equation}\label{q11}
\begin{aligned} 
&\int_{0}^{\infty}v_ndv_n\int_{0}^{2\pi} d\Psi_n\; \boldsymbol{D}_{v_n,\Psi_n}^k \mathcal{F}(v_n;\bar{v}_n,\Psi_n)
\\&=\int_{0}^{\infty}dv_n\; \frac{(-1)^k 4^k k!}{j_n\{2\}^k} \mathcal{F}(v_n;\bar{v}_n)
\\&\;\times\Big((\mathit{L}_k (\frac{v_n^2+\bar{v}_n^2}{j_n\{2\}})+\alpha_k)I_0(\frac{2v_n\bar{v}_n}{j_n\{2\}})+\sum_{l=1}^{k}\beta_{kl}I_l(\frac{2v_n\bar{v}_n}{j_n\{2\}})\Big).
\end{aligned} 
\end{equation}
The radial flow distribution $p(v_n;\bar{v}_n)$ using Eq.(\ref{q11}) can be obtained
\begin{equation}\label{q12}
\begin{aligned} 
&p_q(v_n;\bar{v}_n)
\\&=\int_{0}^{2\pi} d\Psi_n v_n p(v_n;\bar{v}_n,\Psi_n)
\\&\approx\int_{0}^{2\pi} d\Psi_n v_n\left[1+\sum_{k=2}\frac{j_n\{2k\}\boldsymbol{D}_{v_n,\Psi_n}^{k}}{4^{k}(k!)^2}\right]\mathcal{F}(v_n;\bar{v}_n,\Psi_n)\\ &=\mathcal{F}(v_n;\bar{v}_n)\sum_{k=0}^{q}\frac{(-1)^k\gamma_{k}}{k!}\Big[\alpha^{'}_{k}I_0(\frac{2v_n\bar{v}_n}{j_n\{2\}})
+\sum_{l=0}^{k}\beta_{kl}I_l(\frac{2v_n\bar{v}_n}{j_n\{2\}})\Big],
\end{aligned}
\end{equation}
where $\alpha^{'}_{k}\equiv\mathit{L}_k (\frac{v_n^2+\bar{v}_n^2}{j_n\{2\}})+\alpha_k$ and $\gamma_{k}\equiv j_n\{2k\}/j_n\{2\}^k=q_n\{2k\}/q_n\{2\}^k+\cdots$.
Note that we have assumed $\gamma_0=1$ and $\gamma_1=\alpha_0=\beta_{k0}=0$ in Eq.(\ref{q12}). 
The first term ($q=0$) of $p_q(v_n;\bar{v}_n)$ is a Bessel-Gaussian distribution. Other terms are the corrections to the Bessel-Gaussian distribution. 
\begin{figure}[t!]
		\begin{tabular}{c}
			\includegraphics[scale=0.4]{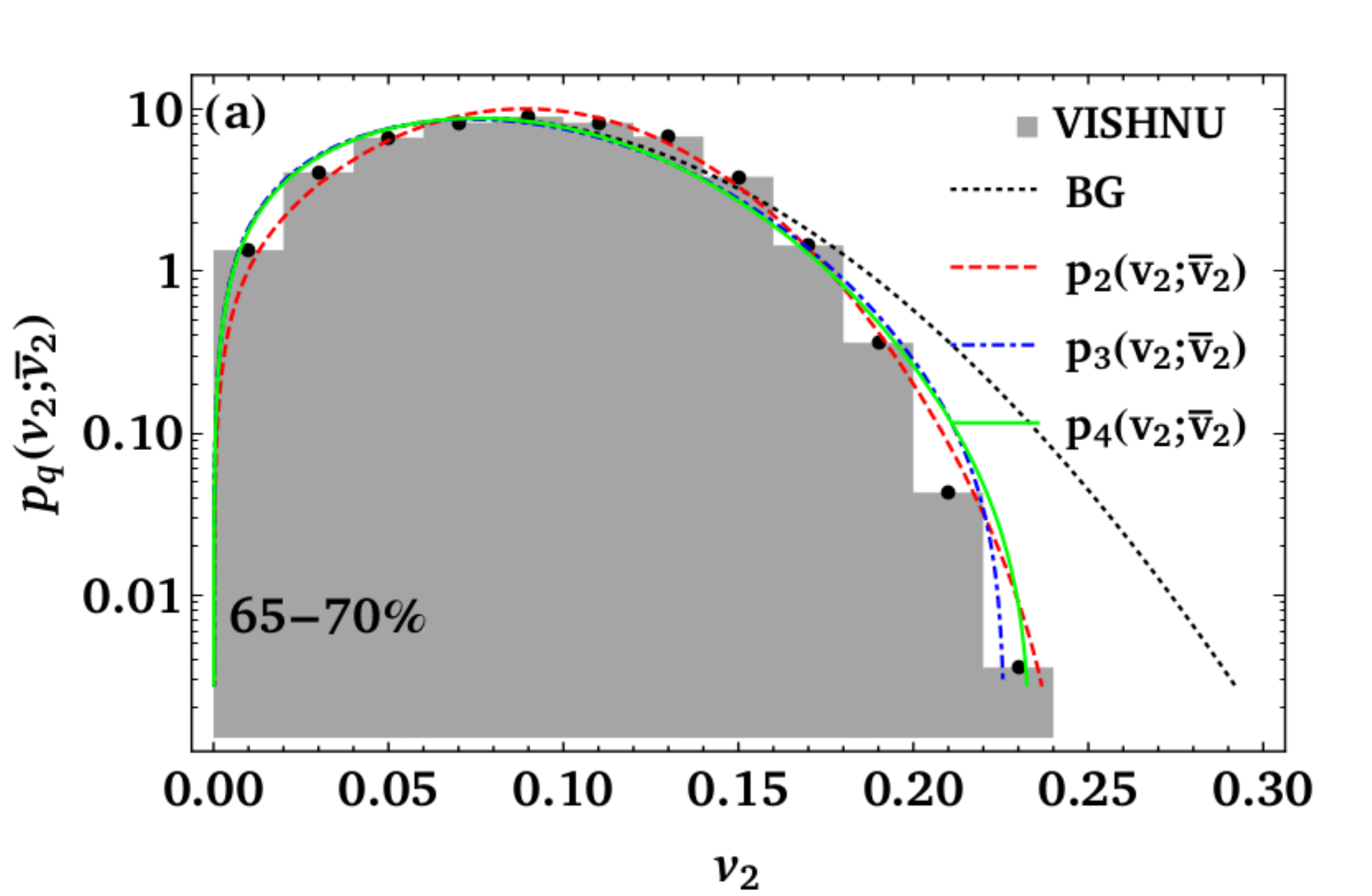}\\
			\includegraphics[scale=0.4]{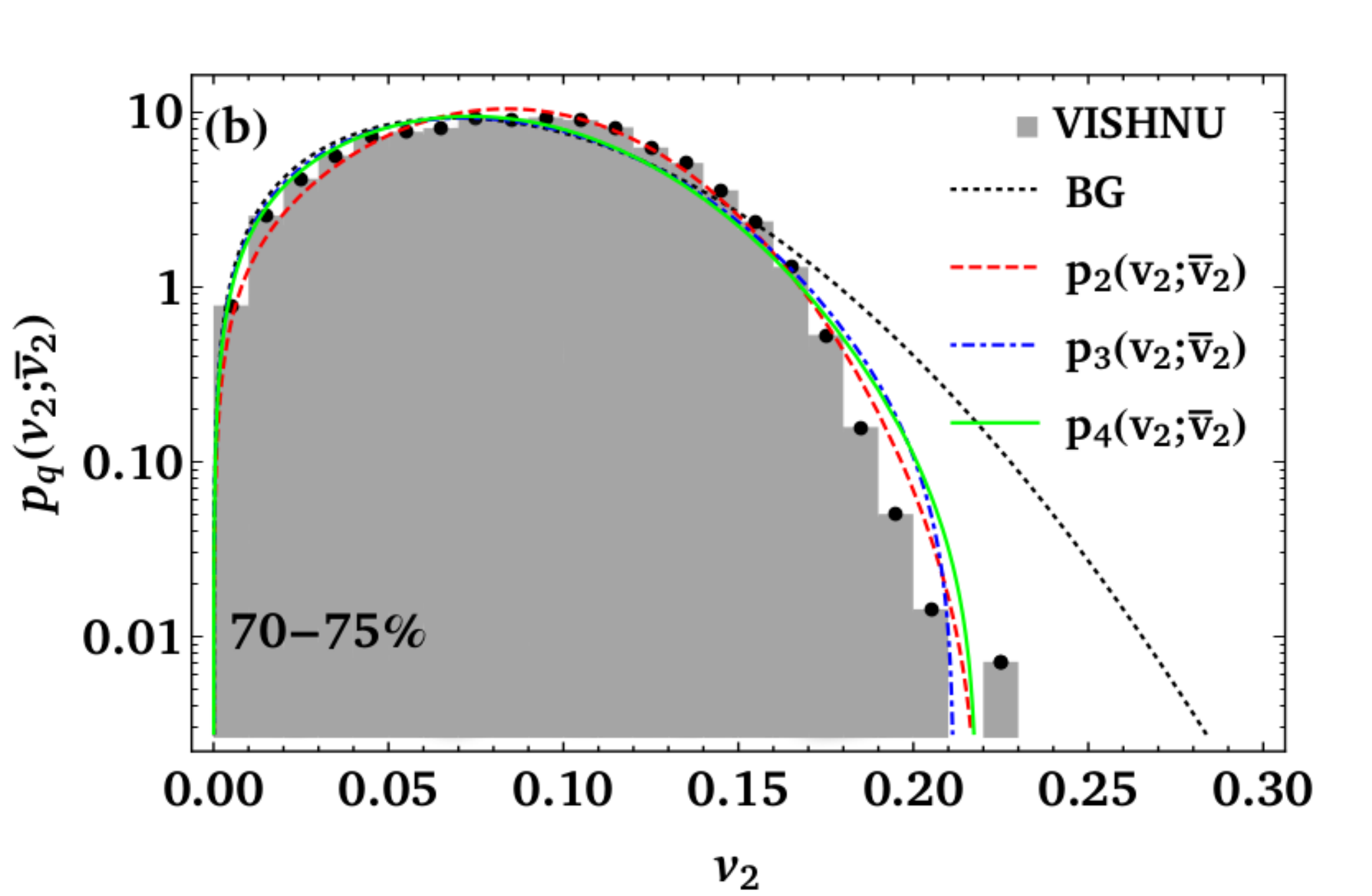} \\
			\includegraphics[scale=0.4]{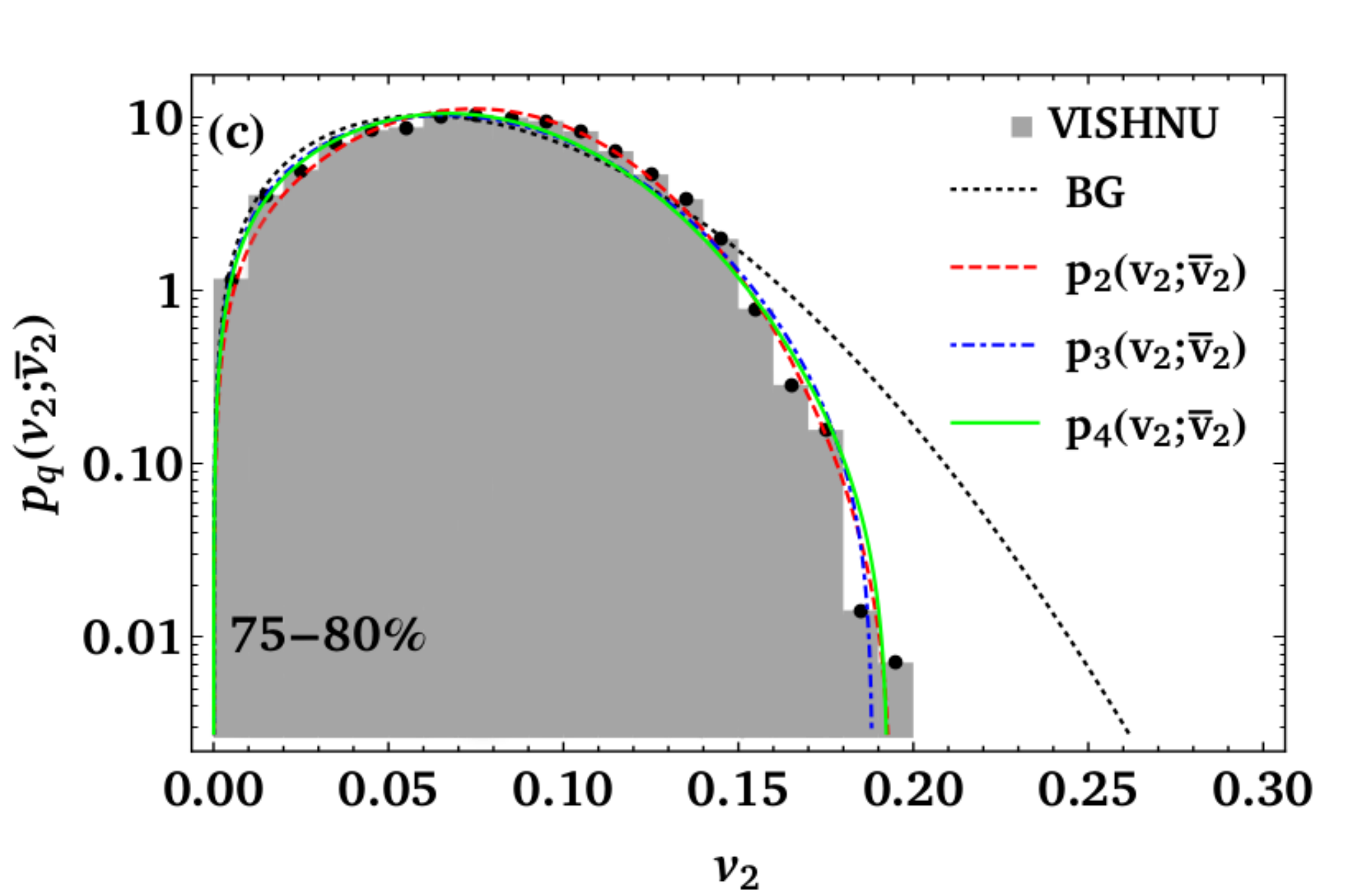}
		\end{tabular}		
		\caption{(Color online) Comparing the obtained flow distribution from iEBE-VISHNU output with different truncations of distribution $p_q(v_n;\bar{v}_n)$ for $q=0,2,3,4$ presented by dotted black, dashed red, dot-dashed blue, solid green lines, respectively.} 
		\label{Check3}
\end{figure}

\begin{figure}[t!]
	\begin{tabular}{c}
		\includegraphics[scale=0.49]{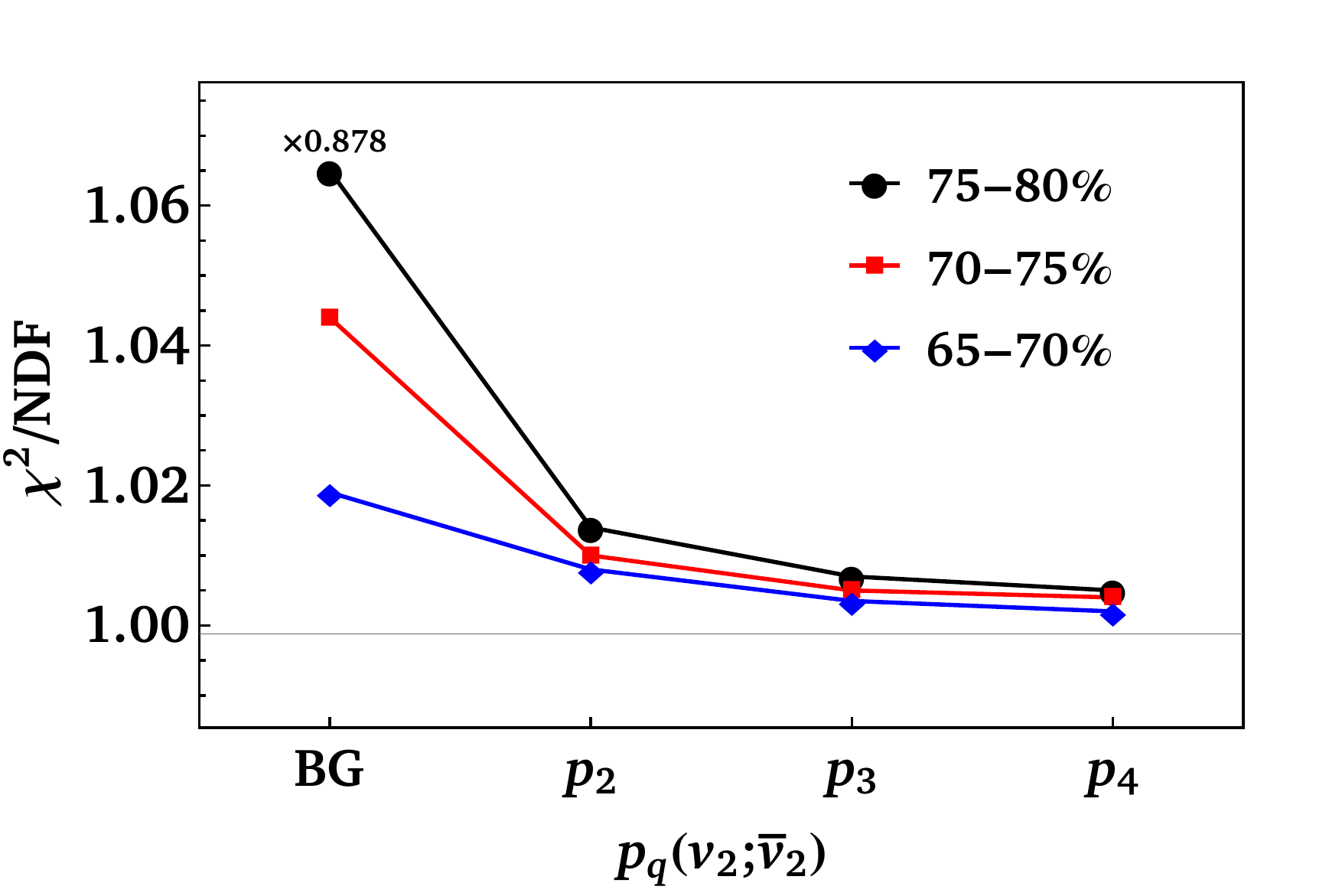}
	\end{tabular}		
	\caption{(Color online) $\chi^2/$NDF values of fitting the Bessel-Gaussian distribution and $p_q(v_n;\bar{v}_n)$ for $q=2,3,4$ to simulation plotted in  $65-70\%$, $70-75\%$, and $75-80\%$ centrality collisions. The corresponding $\chi^2/$NDF value of the Bessel Gaussian distribution multiplied by $0.878$. This is done to increase the resolution in other parts of the plot.} 
	\label{Check6}
\end{figure}

Fig. \ref{Check3} compares the obtained distribution from iEBE-VISHNU with estimated distribution in Eq.(\ref{q12}). In this figure, we investigate different truncations of $p_q(v_n;\bar{v}_n)$ for $q=0,2,3,4$ presented by dotted black, dashed red, dot-dashed blue, solid green lines, respectively. Since the main shortcoming of the Bessel-Gaussian distribution compared with the simulation data are in peripheral collisions, we only show the results in $65-70\%$, $70-75\%$, and $75-80\%$ centrality classes. As demonstrated in this figure, the generated data cannot be described by the black curve, which corresponds to the Bessel-Gaussian distribution. 
Also, studying $\chi^2/$NDF for the Bessel-Gaussian distribution and $p_q(v_n;\bar{v}_n)$ for $q=2$, $3$, and $4$ plotted in Fig. \ref{Check6} for the investigated centralities in Fig. \ref{Check3}. 
As can be seen, the values of $\chi^2/$NDF associated $p_q(v_n;\bar{v}_n)$ are more closer to $1$ comparing with the Bessel Gaussian distribution.  
The results of Fig. \ref{Check3} and \ref{Check6} show that the distribution of elliptic flow is deviated from the Bessel Gaussian distribution. So, the corrections to the Bessel Gaussian distribution becomes important which is described by $p_q(v_n;\bar{v}_n)$.    
\section{Joint Flow Distribution}\label{sec4}
\begin{figure}[t!]
	\begin{tabular}{c}
		\includegraphics[scale=0.49]{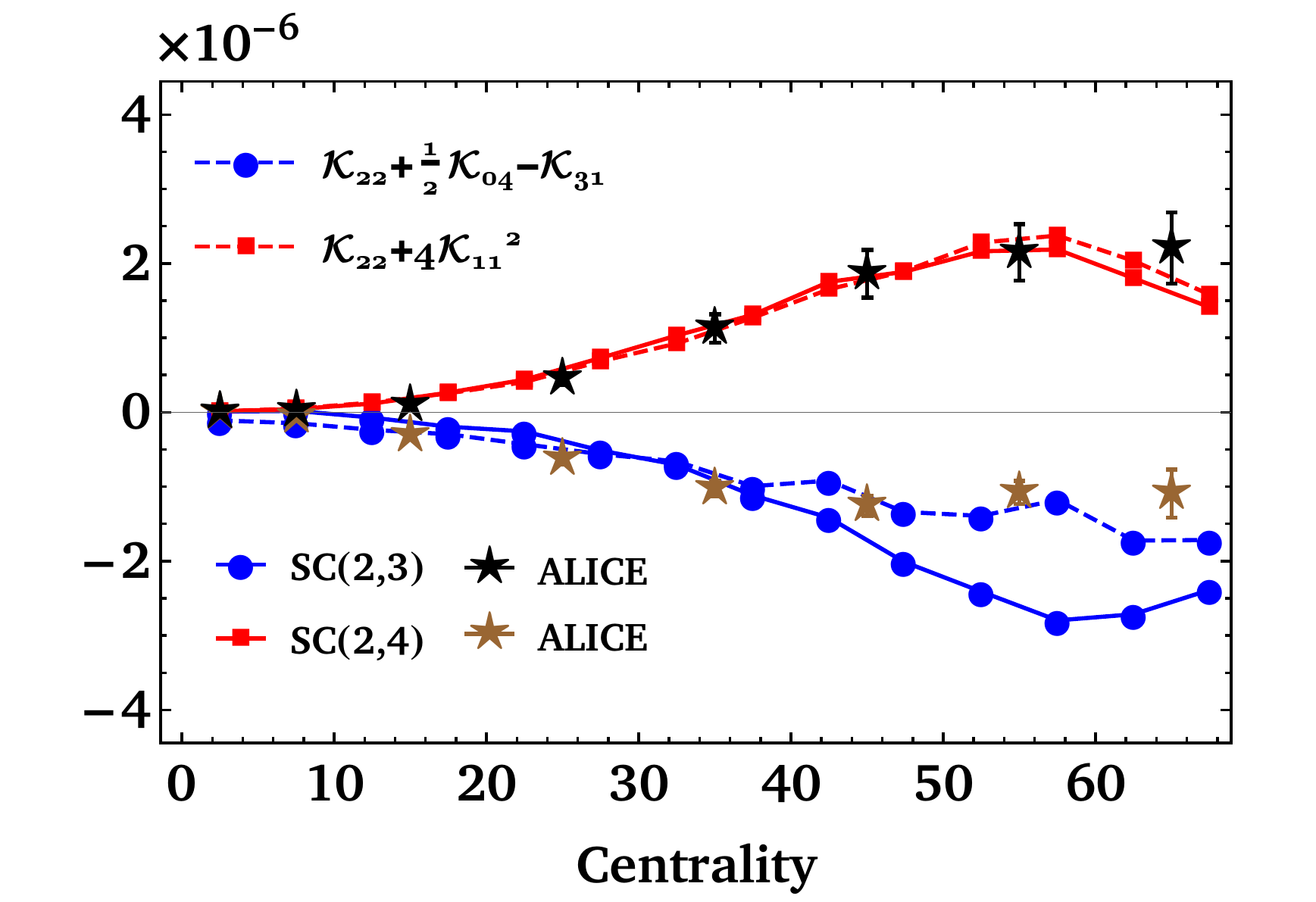}
	\end{tabular}		
	\caption{(Color online) Comparing the combinations of joint cumulants $\K$, $\K_{22}+\frac{1}{2}\K_{04}-\K_{31}$ and $\K_{22}+4\K_{11}^2$, and symmetric cumulants obtained from iEBE-VISHNU with the results of ALICE Collaboration \cite{Aad:2014vba}.} 
	\label{Check4}
\end{figure}

The information of the event-by-event flow fluctuations are encoded in the joint flow harmonic distribution $p(v_1,v_2,...)$, as mentioned in Sec. \ref{sec1}. Therefore, using joint cumulant and moment generating functions, we obtain joint distribution of flow harmonics in this section.
To do this, we consider the relation between joint generating function of moments and cumulants as follows
\begin{equation}\label{q16}
\la e^{W_n\lambda_{n} + W_m\lambda_{m}}\ra=\exp\left[\sum_{k,l=0}\frac{(\lambda_{n})^{k}(\lambda_{m})^l}{k!l!}\K_{kl}\right],
\end{equation}
where $W_n$ and $\K_{nm}$ are shifted flow vectors and joint flow cumulants, respectively. 
The relations between $\K_{kl}$ and moments are
\begin{subequations}
	\begin{eqnarray}
\K_{00}&=&\K_{10}=\K_{01}=0,\\
\K_{11}&=&\la W_n W_m^*\ra=\la v_nv_m\cos(\Psi_1-\Psi_2)\ra-\bar{v}_n\bar{v}_m,\quad\\
\K_{20}&=&\la |W_n|^2\ra=\la v_n^2\ra-\bar{v}_n^2,\\
\K_{02}&=&\la |W_m|^2\ra=\la v_m^2\ra-\bar{v}_m^2,\\
\vdots 
	\end{eqnarray}
\end{subequations}
Note that because the average of sifted flow vector $\la W_n\ra$ is zero the cumulants $\K_{10}$ and $\K_{01}$ are zero for all harmonics. Using Eq.(\ref{eq12}), we can rewrite the cumulants of flow joint distribution in terns of $j_n\{2k\}$ and flow correlations are as following
\begin{subequations}\label{eq18}
	\begin{eqnarray}
	\K_{11}&=&Re[\la V_nV_m'^{*}\ra]-\bar{v}_n\bar{v}_m,\\
	\K_{20}&=&j_n\{2\},\\
	\K_{02}&=&j_m\{2\},\\
	\vdots 
	\end{eqnarray}
\end{subequations}
One way to investigate the event-by-event flow fluctuations is by measuring  the  correlation between the  magnitudes of different flow harmonics using a cumulant analysis. These new observables are commonly known as \textit{Symmetric Cumulants} ($SC$). Recently, ALICE has measured $SC(2,3)$ and $SC(2,4)$ as a function of centrality \cite{Aad:2014vba} at center-of-mass energy per nucleon pair $\sqrt{s}= 2.76$ TeV, with transverse momentum in the range of $0.2<p_T<5\;\text{GeV}$. In this paper, we show that these experimental data can be explained by a combination of joint cumulants $\K$. Fig. \ref{Check4} present a comparison between simulation and experimental data.
It is worth mentioning that using VISHNU output $p_T$ is in the range $0.28<p_T<4\;\text{GeV}$. As can be seen, there is a mismatch between $SC(2,3)$ obtained from simulation and experiment. But the experimental data can be described by combination $\K_{22}+\frac{1}{2}\K_{04}-\K_{31}$. Also, one can find that $SC(2,4)=\K_{22}+4\K_{11}^2$ can explain the ALICE data \footnote{In Ref.\cite{Mordasini:2019hut}, authors have compared $SC(2,3)$ and $SC(2,4)$ obtained from iEBE-VISHNU in three different transverse momentum ranges with the experimental results from ALICE (see Fig.12 in Ref.\cite{Mordasini:2019hut}). They have found that $SC(2,3)$ and $SC(2,4)$ obtained from ALICE data can be explained by $SC(2,3)$ and $SC(2,4)$ obtained from iEBE-VISHNU in different $p_T$ ranges. Instead, we have kept $p_T$ range the same and tried to find the combinations of $\mathcal{K}_{mn}$ to explain the experimental results in a transverse momentum range.}. 
Now, having the joint cumulants  enables us to obtain the joint distribution of flow harmonics. To do this, one should find a form of cumulative characteristic function $G(\lambda_{n},\lambda_{m})$ by expanding it to $k+l=2$, 
\begin{equation}
\begin{aligned}
\exp\Big(G&(\lambda_{n},\lambda_{m})\Big)\\&=\exp\left[\sum_{k+l\geq 3}\tilde{\K}_{kl} (\lambda_{n})^{k}(\lambda_{m})^{k}\right]\\&\hspace*{.7cm}\times\exp\left[\lambda_{n}^2\tilde{\K}_{20}+\lambda_{m}^2\tilde{\K}_{02}+\lambda_{n}\lambda_{m}\tilde{\K}_{11} \right]\\
&=\exp\left[\sum_{k+l\geq 3}\tilde{\K}_{kl}(\lambda_{n})^{k}(\lambda_{m})^{l}\right]\mathcal{N}(\lambda_{n},\lambda_{m}).
\end{aligned}
\end{equation}
where the standard joint cumulants $\tilde{\K}_{mn}$ are $\K_{mn}/(m!n!)$.
Applying Fourier transforming to both sides of Eq.(\ref{q16}), we get
\begin{equation}
\begin{aligned}
\int \boldsymbol{d}W_n\boldsymbol{d}W_m&\;\mathcal{P}(W_n,W_m)e^{W_n\lambda_{n}+W_m\lambda_{m}}\\&=\exp\left[\sum_{k+l\geq 3}\tilde{\K}_{kl}(\lambda_{n})^{k}(\lambda_{m})^{l}\right]\mathcal{N}(\lambda_{n},\lambda_{m}),
\end{aligned}
\end{equation} 
where $\boldsymbol{d}W_n=dw_{n,x}dw_{n,y}$. Eventually, we find the joint distribution $\mathcal{P}(W_n,W_m)$ as
\begin{equation}\label{q24} 
\begin{aligned}
&\mathcal{P}(W_n,W_m)
\\&=\frac{1}{2\pi\Delta}\exp\left[\sum_{k+l\geq 3}\tilde{\K}_{kl} (\partial_{n})^{k}(\partial_{m})^{l}\right]\\&\hspace*{0cm}\times\exp\left[-\frac{\tilde{\K}_{02}w_n^2+\tilde{\K}_{20}w_m^2-\tilde{\K}_{11}(w_{n,x}w_{m,x}+w_{n,y}w_{m,y})}{\Delta^2} \right]\\
&\approx\left[1+\sum_{k+l\geq 3}\tilde{\K}_{kl} (\partial_{n})^{k}(\partial_{m})^{l}\right]\mathcal{N}(W_n,W_m)\\
\end{aligned}
\end{equation}
where $\Delta$ defined $(4\tilde{\K}_{20}\tilde{\K}_{02}-\tilde{\K}_{11}^2)^{1/2}$ or in the simplified case $(j_n\{2\}j_m\{2\}-Re[\la V_nV_m'^*\ra^2])^{1/2}$.
Note that if we only consider $\mathcal{N}(W_n,W_m)$ as the first term of $\mathcal{P}(W_n,W_m)$ and compare it with bivariate normal distribution\footnote{In 2D, the probability density function of a vector $[x'y']$ is 
\begin{align*}
f(x',y')=&\frac{1}{2\pi\sigma_m\sigma_n\sqrt{1-\rho^2}}
\\&\times\exp\Big(-\frac{1}{2(1-\rho^2)}\Big[\frac{x'^2}{\sigma_{x'}^2}+\frac{y'^2}{\sigma_{y'}^2}-\frac{2\rho x'y'}{\sigma_{x'}\sigma_{y'}}\Big]\Big).
\end{align*}
where $\sigma$ and $\rho$ are the standard deviation and the Pearson correlation, respectively.}, 
we find that
\begin{subequations}\label{qq19}
	\begin{eqnarray}
	\sigma_{n}^2&=&2\tilde{\K}_{20}=\la |W_n|^2\ra,\\
	\sigma_{m}^2&=&2\tilde{\K}_{20}=\la|W_m|^2\ra,\\
	\rho_{nm}&=&\frac{\tilde{\K}_{11}}{2\sqrt{\tilde{\K}_{20}\tilde{\K}_{02}}}=\frac{Re[\la W_nW_m'^*\ra]}{\sqrt{\la|W_n|^2\ra\la|W_m|^2\ra}}.
	\end{eqnarray}
\end{subequations} 
These results show that the general joint distribution of flow vectors can be obtained,
\begin{equation}\label{q26}
\begin{aligned}
\mathcal{P}&(W_1,W_2,...,W_n)\\&\approx\left[1+\sum_{k_1+...+k_n\geq 3}\tilde{\K}_{k1...kn}(\partial_1)^{k_1}\cdots(\partial_n)^{k_n}\right]
\\&\hspace*{4cm}\times\mathcal{N}(W_1,W_2,...,W_n),
\end{aligned} 
\end{equation}  
by defining the joint cumulant and moment generating function relation, 
\begin{equation}\label{q27}
\begin{aligned}
\la &e^{W_1\lambda_{1}+...+ W_n\lambda_{n}}\ra\\&\hspace*{1cm}=\exp\left[\sum_{k_1,..,k_n=0}\frac{(\lambda_{1})^{k_1}\cdots(\lambda_{n})^{k_n}}{k_1!...k_n!}\K_{k1...kn}\right],
\end{aligned}
\end{equation}
where $\K_{k1...kn}$ are the generalized joint cumulants, and the first cumulant, $\mathcal{K}_{0...0}$, is equal to zero by considering the normalization condition of the probability distribution. 
Note that the distribution $\mathcal{N}(W_1,W_2,...,W_n)$ in Eq.(\ref{q26}) is a generalization of the one-dimensional normal distribution to higher dimensions which is dubbed as the joint normal distribution. But it should be noted that since the souls of the multivariate normal distribution and the distribution $\mathcal{N}(W_1,W_2,...,W_n)$ in Eq.(\ref{q27}) are different, we have $\int\boldsymbol{d}W_1\cdots\boldsymbol{d}W_n\;\mathcal{N}(W_1,W_2,...,W_n)\neq1$. So, in the following, we use normalized kernel $\mathcal{N}(W_1,W_2,...,W_n)$
to find the joint distribution of flow magnitudes.

\begin{figure}[t!]
	\begin{tabular}{c}
		\includegraphics[scale=0.49]{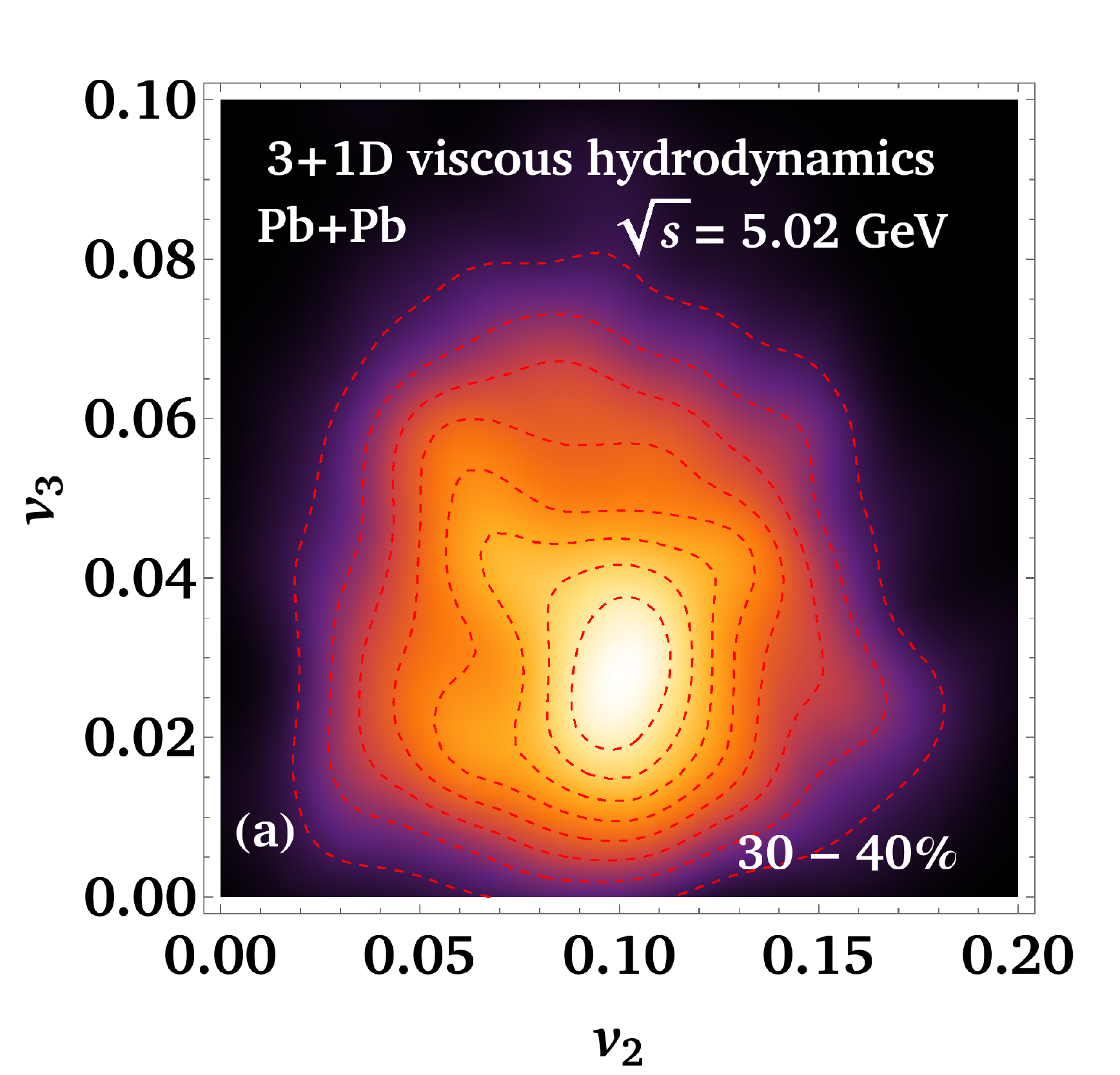}\\
		\includegraphics[scale=0.49]{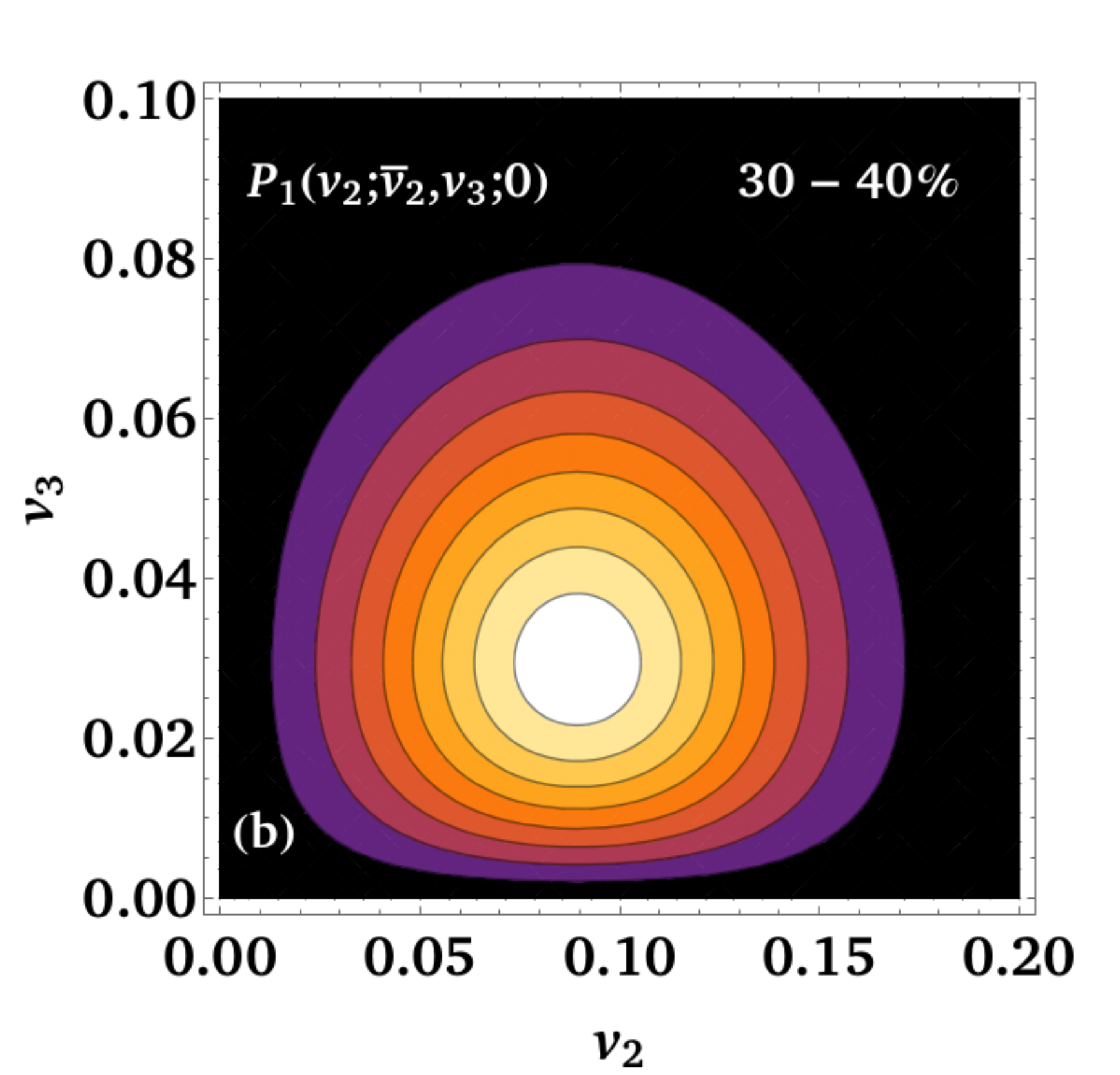}
	\end{tabular}		
	\caption{(Color online) Comparing the smooth density histogram of $v_2$ and $v_3$ obtained from $3+1$D viscous hydrodynamic simulation \cite{Bozek:2009dw} and the contour plot of the first term of distribution $\mathcal{P}(v_2;\bar{v}_2,v_3;0)$ in $30-40\%$ centrality class.} 
	\label{Check5}
\end{figure}

Let us return to the computation of the joint radial distribution of two flow harmonics using Eq.(\ref{q24}). In Ref. \cite{Voloshin:1994mz}, a technique has been introduced that enables us to study the correlations between any rapidity windows and any harmonics. In this technique denoting the relative angle $\Phi=\Psi_m-\Psi_n$\footnote{Note that the angles $\Phi$ and $\Psi$ are in the range of $[0,\pi]$ and $[0,2\pi]$, respectively}, and averaging over reaction plane angle, the joint radial flow distribution is obtained as
\begin{equation}\label{q21} 
\begin{aligned}
\int &dv_ndv_m\mathcal{P}(v_n;\bar{v}_n,v_m;\bar{v}_m)
\\
&\equiv\int v_ndv_n v_mdv_m\int \frac{\boldsymbol{d}\mathcal{P}(W_n,W_m)}{d\Psi_m d\Psi_n}d\Psi_m d\Psi_n d\Phi
\\&\hspace*{4.5cm}\times\delta(\Phi-\Psi_m+\Psi_n).
\end{aligned}
\end{equation}
To study the joint radial flow distribution, we consider the first term in Eq.(\ref{q24}) for simplicity. Inserting it in Eq.(\ref{q21}) we obtain the joint flow distribution\footnote{Here to normalize distribution $\mathcal{N}(W_n,W_m)$, we assume $\Delta\to\Delta/\sqrt{2}$.}:
\begin{equation}\label{q22} 
\begin{aligned}
\int& dv_ndv_m\mathcal{P}_1(v_n;\bar{v}_n,v_m;\bar{v}_m)
\\&=\int dv_ndv_md\Phi\mathcal{\chi}_{mn}e^{\zeta_3\cos\Phi}I_0\left(\sqrt{\zeta_1^2+\zeta_2^2+2\zeta_1\zeta_2\cos(\Phi)}\right)
\end{aligned}
\end{equation}
where
\begin{subequations}
	\begin{eqnarray}
\mathcal{\chi}_{mn}&\equiv&\frac{4v_nv_m}{\pi\Delta^2}\exp\left[-\frac{v_n^2+\bar{v}_n^2}{\Delta^2/2\tilde{\K}_{02}}-\frac{v_m^2+\bar{v}_m^2}{\Delta^2/2\tilde{\K}_{20}}+\frac{\bar{v}_n\bar{v}_m}{\Delta^2/2\tilde{\K}_{11}}\right],\quad\quad\\
\zeta_1&\equiv& v_n\Big(\frac{2\bar{v}_n}{\Delta^2/2\tilde{\K}_{02}}-\frac{\bar{v}_m}{\Delta^2/2\tilde{\K}_{11}}\Big),\\ 
\zeta_2&\equiv& v_m\Big(\frac{2\bar{v}_m}{\Delta^2/2\tilde{\K}_{20}}-\frac{\bar{v}_n}{\Delta^2/2\tilde{\K}_{11}}\Big),\\
\zeta_3&\equiv& \frac{\bar{v}_n\bar{v}_m}{\Delta^2/2\tilde{\K}_{11}}.
	\end{eqnarray}
\end{subequations}
Note that Eq.(\ref{q22}) is the first approximation of the radial joint distribution of any two flow harmonics. To study the distribution $\mathcal{P}(v_n;\bar{v}_n,v_m;\bar{v}_m)$, we investigate it for $v_2$ and $v_3$. In this case, since the triangular flow distribution is rotationally symmetric, $\bar{v}_3$ is zero. Also, as mentioned above, $\Delta_{v_2,v_3}^2$ is $j_2\{2\}j_3\{2\}-Re[\la V_2V_3^*\ra^2]$. If we check $\Delta_{v_2,v_3}^2$, we find that the term $Re[\la V_2V_3^*\ra^2]$ is very small and negligible against $j_2\{2\}j_3\{2\}$. So, we can write $\Delta_{v_2,v_3}^2\simeq j_2\{2\}j_3\{2\}$. It should be noticed that the contribution of $Re[\la V_2V_3^*\ra^2]$ in $\zeta_i$ is non-negligible, because it is in the numerator. Concerning these variables, the joint distribution of second and third harmonics can be rewritten
\begin{equation}\label{q30}
\begin{aligned}
\mathcal{P}_1(v_2;\bar{v}_2,v_3;0)&=\frac{4v_2v_3}{\pi j_2\{2\}j_3\{2\}}\exp\left[-\frac{v_2^2+\bar{v}_2^2}{j_2\{2\}}-\frac{v_3^2}{j_3\{2\}}\right]
\\&\times\int d\Phi I_0\left(\sqrt{\gamma_1^2+\gamma_2^2+2\gamma_1\gamma_2\cos(\Phi)}\right),
\end{aligned}
\end{equation}
Where 
\begin{equation}
\gamma_1\equiv \frac{2v_n\bar{v}_n}{\K_{20}}, \quad 
\gamma_2\equiv -\frac{v_m\bar{v}_n}{\K_{20}\K_{02}/2\K_{11}}.
\end{equation}
Eq.(\ref{q30}) shows that because there is a non-negligible correlation between $V_2$ and $V_3$ the first approximation of joint distribution $\mathcal{P}(v_2;\bar{v}_2,v_3;0)$ cannot be written as $p_0(v_3;0)p_0(v_2;\bar{v}_2)$. Note that $p_0(v_n;\bar{v}_n)$ is the first truncation of the distribution Eq.(\ref{q12}) which is called the Bessel-Gaussian distribution. However, we checked that if one consider $\mathcal{P}(v_2;\bar{v}_2,v_3;0)=p_0(v_3;0)p_0(v_2;\bar{v}_2)$, its results and the results of Eq.(\ref{q30}) approximately are the same.     
Fig.(\ref{Check5}) present the smooth density histogram of $v_2$ and $v_3$, which is obtained by using the results of the event-by-event 3 + 1D viscous hydrodynamics at center-of-mass energy per nucleon pair $\sqrt{s}=5.02$ TeV \cite{Bozek:2009dw}. These data are obtained with the wounded-quark initial conditions, and the contour plot of the first term of distribution $\mathcal{P}(v_2;\bar{v}_2,v_3;0)$ in $30-40\%$ centrality. As can be seen in this figure, the results show that there is a decent agreement between theory and simulation data. To find the best estimation, one have to insert the complete form of Eq.(\ref{q24}) in Eq.(\ref{q21}).

\section{conclusion}\label{con}
In this paper, we employed the relation between joint cumulant and moment generating function of $v_{n,x}$ and $v_{n,y}$ to relate the radial flow distribution to cumulants by using the standard method of finding Gram-Charlier series. We have found a general flow distribution in Eq.(\ref{q12}) by using Fourier transformation both sides of Eq.(\ref{q9}). It is an expansion around Bessel-Gaussian distribution where the coefficients of the expansion have been written in terms of shifted cumulants $j_n\{2k\}$.
We have shown that $p(v_n;\bar{v}_n)$ can explain the generated data in the peripheral collisions, by assuming $\bar{v}_n\neq0$ for even harmonics. Our results indicate a significant improvement over the Bessel-Gaussian distribution. Also, we have obtained the odd flow distribution which has been found in Refs. \cite{navid} and \cite{hadi} by setting $\bar{v}_n=0$.
The shifted cumulants $j_n\{2k\}$ were written in terms of moments $\la w_n^k\ra$ where $w_n$ is the magnitude of the shifted flow vector. If one assumes $\bar{v}_n=0$, the cumulants $j_n\{2k\}$ would be 2k-particle correlation functions $c_n\{2k\}$ which can be observed experimentally. Also, we have shown that the shifted cumulants $j_n\{2k\}$, which is obtained from the relation between joint cumulant and moment generating functions, have more information than the cumulants $q_n$ found in previous works. In the final step, we have studied
the joint distribution of flow harmonics and presented a general form for $\mathcal{P}(W_1,W_2,...,W_n)$. To do this, we introduced new observables $\K_{nm}$ and showed that the experimental data for symmetric cumulants $SC(2,3)$ and $SC(2,4)$ can be explained by combinations of these observables. So, we think that the cumulants $\K_{mn}$ can be interesting observables for experimentalists. We also obtained the joint radial distribution of the two flow harmonics, and showed that the first terms of this distribution for $v_2$ and $v_3$ can justify the simulation data. Investigating joint distributions of other flow harmonics is left to future studies.     
\section{Acknowledgment}
We would like to thank  Jean-Yves Ollitrault and Fatemeh Elahi for useful discussions and comments, as well as Hessamadin Arfaei for all his supports.

\appendix
\section{Odd Flow Distribution}\label{app0} 
As mentioned in Sec. \ref{sec2}, in Eq.(8) only terms with $k=l$ are non-zero. Moreover, the Eq.(8) can be written as
\begin{equation}
G(\lambda_x,\lambda_y)=\la e^{v_{n,x}\lambda_x + v_{n,y}\lambda_y}\ra
\exp\left[\sum_{k=1}\frac{(\lambda_x^2+\lambda_y^2)^{k}}{2^{(2k)}(k!)^2}c_n\{2k\}\right].
\end{equation} 
Since we would like to use the Gram-Chalier series with normal kernel, we can rewrite $G(\lambda_x,\lambda_y)$ as following
\begin{equation} 
G(\lambda_x,\lambda_y)=\exp\left[\sum_{k=2}\frac{(\lambda_x^2+\lambda_y^2)^{k}}{2^{(2k)}(k!)^2}c_n\{2k\}\right]e^{\frac{\lambda_x^2+\lambda_y^2}{4}c_n\{2\}}.
\end{equation} 
We can find the odd flow distribution $p_{odd}(v_{n,x},v_{n,y})$ by using inverse Fourier transform of $G(\lambda_x,\lambda_y)$,
{\small\begin{equation}
	\begin{aligned}
p_{odd}&(v_{n,x},v_{n,y})
\\&=\frac{1}{(2\pi)^2}\int_{-\infty}^{\infty}\int_{-\infty}^{\infty}d\lambda_xd\lambda_y e^{-i(v_{n,x}\lambda_x + v_{n,y}\lambda_y)}G(\lambda_x,\lambda_y)\\
&=\frac{1}{(2\pi)^2}\int_{-\infty}^{\infty}\int_{-\infty}^{\infty}d\lambda_xd\lambda_y e^{-i(v_{n,x}\lambda_x + v_{n,y}\lambda_y)}
\\& \times\exp\left[\sum_{k=2}\frac{((-i\lambda_x)^2+(-i\lambda_y)^2)^{k}}{2^{(2k)}(k!)^2}c_n\{2k\}\right]e^{-\frac{\lambda_x^2+\lambda_y^2}{4}c_n\{2\}}.\\
\end{aligned}\end{equation}}
Also, we mentioned that $(it)^n G_N(t)$ is the characteristic function of $(-\boldsymbol{D})^n G_N(x)$, we have
{\small\begin{equation}\begin{aligned}
p_{odd}&(v_{n,x},v_{n,y})
\\&=\frac{1}{(2\pi)^2}\exp\left[\sum_{k=2}\frac{(\partial_x^2+\partial_y^2)^{k}}{2^{(2k)}(k!)^2}c_n\{2k\}\right]
\\&\times\int_{-\infty}^{\infty}\int_{-\infty}^{\infty}d\lambda_xd\lambda_y e^{-i(v_{n,x}\lambda_x + v_{n,y}\lambda_y)}e^{-\frac{\lambda_x^2+\lambda_y^2}{4}c_n\{2\}}
\\&=\frac{1}{(2\pi)^2}\exp\left[\sum_{k=2}\frac{(\partial_x^2+\partial_y^2)^{k}}{2^{(2k)}(k!)^2}c_n\{2k\}\right]
\\&\times\left[\frac{4\pi}{c_n\{2\}}e^{-\frac{v_{n,x}^2+v_{n,y}^2}{c_n\{2\}}}\right]
\\&=\exp\left[\sum_{k=2}\frac{c_n\{2k\}(\partial_x^2+\partial_y^2)^{k}}{2^{2k}(k!)^2}\right]\left[\frac{1}{\pi c_n\{2\}}e^{\frac{-v_{n,x}^2-v_{n,y}^2}{ c_n\{2\}}}\right].
\end{aligned}\end{equation}}

\section{General Form of Radial Derivatives}\label{app1} 
If we differentiate 1D normal distribution, $\mathcal{N}(r)=e^{-\frac{r^2}{a}}/(\sqrt{\pi a})$ with $a=2\sigma^2$, $n$ times, the result of each time is approximately a Laguerre polynomial,
\begin{equation} 
\begin{aligned}
k=1&:D_r^1 \mathcal{N}(r) = D_r (\mathcal{N}(r) \mathit{L}_0 (\frac{r^2}{a})),
\\k=2&:D_r^2 \mathcal{N}(r) \approx D_r (\mathcal{N}(r) \mathit{L}_1 (\frac{r^2}{a})),
\\k=3&:D_r^3 \mathcal{N}(r) \approx D_r (\mathcal{N}(r) \mathit{L}_2 (\frac{r^2}{a})),
\\k=4&:D_r^4 \mathcal{N}(r) \approx D_r (\mathcal{N}(r) \mathit{L}_3 (\frac{r^2}{a})),
\\\vdots
\\k=n&:D_r^n (e^{-\frac{r^2}{a}}) \approx D_r (\mathcal{N}(r) \mathit{L}_n (\frac{r^2}{a})).
\end{aligned}
\end{equation}
Note that the radial derivative is $D_{r}=\partial_r^2+(1/r)\partial_r$. 
Using above derivative in Eq.(\ref{q7}), we can rewrite the $p(v_n)$ as follows   
\begin{equation}\label{aq1}
\begin{aligned}
p_{odd}&(v_n) = \frac{2v_n}{c_n\{2\}}\Big[e^{-\frac{v_n^2}{c_n\{2\}}}+\sum_{k=2} \frac{ c_n \{2k\}}{4^k (k!)^2}
\\&\times((-\frac{4}{c_n\{2\}})^{k-1}(k-1)!)
D_{v_n} (e^{-\frac{v_n^2}{c_n\{2\}}} \mathit{L}_{k-1} (\frac{v_n^2}{c_n\{2\}}))\Big].
\end{aligned} 
\end{equation}
If we differentiate $\mathcal{N}(r)\mathit{L}_{k-1} (\frac{r^2}{a})$ in the radial direction,
\begin{equation}\label{aq2}
D_r (\mathcal{N}(r) \mathit{L}_{k-1} (\frac{r^2}{a}))=-\frac{4n}{a}\mathcal{N}(r)\mathit{L}_{k} (\frac{r^2}{a}),
\end{equation}
and then replace it in Eq.(\ref{aq1}), in the result we have the distribution of odd flow harmonics as
\begin{equation}\label{aq3}
\begin{aligned} 
p_{odd}(v_n) &= (\frac{2v_n}{c_n\{2\}})e^{-\frac{v_n^2}{c_n\{2\}}}
\\&\times\Big[1+\sum_{k=2} \frac{(-1)^k c_n\{2k\}}{k!\;c_n\{2\}^k} \mathit{L}_{k} (v_n^2/c_n\{2\})\Big].
\end{aligned}
\end{equation}

\section{Two dimensional derivatives}\label{app2} 
As mentioned in Sec. \ref{sec3}, to find a general flow distribution we can have 
\begin{equation}\label{bq10}
p(v_{n,x},v_{n,y})=\exp\left[\sum_{k=2}\frac{j_n\{2k\}\boldsymbol{D}^{k}}{4^{k}(k!)^2}\right]\mathcal{F}(v_{n,x},v_{n,y}),
\end{equation}
by using Eq.(\ref{q9}) and considering the relation
\begin{equation}
\begin{aligned}
&\int \boldsymbol{D}\boldsymbol{\lambda}\;(\lambda_x^2+\lambda_y^2)^ke^{-i(v_{n,x}-\bar{v}_n)\lambda_x -i v_{n,y}\lambda_y}\times e^{-(\lambda_x^2+\lambda_y^2)j_n\{2\}/4}
\\&= \boldsymbol{D}^k\int \boldsymbol{D}\boldsymbol{\lambda}\;e^{-i(v_{n,x}-\bar{v}_n)\lambda_x -i v_{n,y}\lambda_y}\times e^{-(\lambda_x^2+\lambda_y^2)j_n\{2\}/4}, 
\end{aligned}
\end{equation} 
where $\boldsymbol{D}\boldsymbol{\lambda}=d\lambda_xd\lambda_y$.
If we evaluate the derivative $\boldsymbol{D}$ for $k=1,2,..,k$ in Eq.(\ref{bq10}), we have
\begin{equation}\label{bq3}
\begin{aligned}
k=1:\boldsymbol{D}^1 &\mathcal{F}(v_{n,x},v_{n,y})=
\\& -\frac{4}{j_n\{2\}} \mathcal{F}(v_{n,x},v_{n,y}) \mathit{L}_1 (\frac{(v_{n,x}-\bar{v}_n)^2+v_{n,y}^2}{j_n\{2\}}),
\\&\vdots
\\k=n:\boldsymbol{D}^n &\mathcal{F}(v_{n,x},v_{n,y})= \\&\frac{(-1)^n 4^n n!}{j_n\{2\}^n}\mathcal{F}(v_{n,x},v_{n,y})\mathit{L}_n (\frac{(v_{n,x}-\bar{v}_n)^2+v_{n,y}^2}{j_n\{2\}}).
\end{aligned}
\end{equation} 
Note that the calculations of Eq.(\ref{bq3}) are obtained by using Cartesian partial derivatives. To find radial flow distribution we have to integrate over azimuthal angle. Therefore, it is better to write down in polar coordinates,
\begin{equation}
\begin{aligned}
k=1:&\boldsymbol{D}_{v_n,\Psi_n}^1 \mathcal{F}(v_n;\bar{v}_n,\Psi_n)=\\&-\frac{4}{a} \mathcal{F}(v_n;\bar{v}_n,\Psi_n)\Big(\mathit{L}_1 (\frac{v_n^2+\bar{v}_n^2}{j_n\{2\}})+A_1+B_1\Big),
\\&\vdots
\\k=n:&\boldsymbol{D}_{v_n,\Psi_n}^n \mathcal{F}(v_n;\bar{v}_n,\Psi_n)= \\&\frac{(-1)^n 4^n n!}{a^n} \mathcal{F}(v_n;\bar{v}_n,\Psi_n)\Big(\mathit{L}_n (\frac{v_n^2+\bar{v}_n^2}{j_n\{2\}})+A_n+B_n\Big),
\end{aligned} 
\end{equation}
where $A_k$ and $B_k$ are
\begin{align*}
A_1&=0,\\
B_1&=\frac{2v_n\bar{v}_n}{j_n\{2\}}\cos\Psi_n,\\
&\\
A_2&=\frac{v_n^2\bar{v}_n^2}{j_n\{2\}^2},\\
B_2&=\frac{2v_n\bar{v}_n}{j_n\{2\}}\Big(2\mathit{L}_1 (\frac{v_n^2+\bar{v}_n^2}{2j_n\{2\}})\Big)\cos\Psi_n+\frac{v_n^2\bar{v}_n^2}{j_n\{2\}^2}\cos 2\Psi_n,\\
\end{align*}
\begin{equation} 
\begin{aligned}
A_3&=\frac{v_n^2\bar{v}_n^2}{j_n\{2\}^2}\Big(3\mathit{L}_1 (\frac{v_n^2+\bar{v}_n^2}{3j_n\{2\}})\Big),\\
B_3&=\frac{2v_n\bar{v}_n}{j_n\{2\}}\Big(3\mathit{L}_2 (\frac{v_n^2+\bar{v}_n^2}{2j_n\{2\}})+\frac{1}{8j_n\{2\}^2}(v_n^4+6v_n^2\bar{v}_n^2+\bar{v}_n^4)\Big)\cos\Psi_n\\&+\frac{v_n^2\bar{v}_n^2}{j_n\{2\}^2}\Big(3\mathit{L}_1(\frac{v_n^2+\bar{v}_n^2}{3j_n\{2\}})\Big)\cos 2\Psi_n+\frac{v_n^3\bar{v}_n^3}{3j_n\{2\}^3}\cos 3\Psi_n,\\
\vdots\\
\\A_k&=\alpha_k,
\\B_k&=\sum_{l=1}^{k}\beta_{kl}\cos l\Psi_n.
\end{aligned}
\end{equation}


\begin{thebibliography}{}
	\bibitem{Ackermann:2000tr} 
	K.~H.~Ackermann {\it et al.} [STAR Collaboration],
	Phys.\ Rev.\ Lett.\  {\bf 86}, 402 (2001)
	[nucl-ex/0009011].
	
	
	
	\bibitem{Lacey:2001va} 
	R.~A.~Lacey [PHENIX Collaboration],
	Nucl.\ Phys.\ A {\bf 698}, 559 (2002)
	[nucl-ex/0105003].
	
	
	
	\bibitem{Park:2001gm} 
	I.~C.~Park {\it et al.} [PHOBOS Collaboration],
	Nucl.\ Phys.\ A {\bf 698}, 564 (2002)
	[nucl-ex/0105015].
	
	
	
	\bibitem{Aamodt:2010pa} 
	K.~Aamodt {\it et al.} [ALICE Collaboration],
	Phys.\ Rev.\ Lett.\  {\bf 105}, 252302 (2010)
	[arXiv:1011.3914 [nucl-ex]].
	
		
	\bibitem{ALICE:2011ab} 
	K.~Aamodt {\it et al.} [ALICE Collaboration],
	Phys.\ Rev.\ Lett.\  {\bf 107}, 032301 (2011)
	[arXiv:1105.3865 [nucl-ex]].
	
	\bibitem{Chatrchyan:2012ta} 
	S.~Chatrchyan {\it et al.} [CMS Collaboration],
	Phys.\ Rev.\ C {\bf 87}, no. 1, 014902 (2013)
	[arXiv:1204.1409 [nucl-ex]].
	
	
	
	\bibitem{ATLAS:2011ah} 
	G.~Aad {\it et al.} [ATLAS Collaboration],
	Phys.\ Lett.\ B {\bf 707}, 330 (2012)
	[arXiv:1108.6018 [hep-ex]].
	
	
	\bibitem{Aad:2014vba} 
	G.~Aad {\it et al.} [ATLAS Collaboration],
	Eur.\ Phys.\ J.\ C {\bf 74}, no. 11, 3157 (2014)
	[arXiv:1408.4342 [hep-ex]].
	
\bibitem{Poskanzer:1998yz} 
A.~M.~Poskanzer and S.~A.~Voloshin,
Phys.\ Rev.\ C {\bf 58}, 1671 (1998)
doi:10.1103/PhysRevC.58.1671
[nucl-ex/9805001].



\bibitem{Bhalerao:2003yq} 
R.~S.~Bhalerao, N.~Borghini and J.~Y.~Ollitrault,
Phys.\ Lett.\ B {\bf 580}, 157 (2004)
[nucl-th/0307018].

\bibitem{Bhalerao:2003xf} 
R.~S.~Bhalerao, N.~Borghini and J.~Y.~Ollitrault,
Nucl.\ Phys.\ A {\bf 727}, 373 (2003)
[nucl-th/0310016].

\bibitem{Borghini:2000sa}
N.~Borghini, P.~M.~Dinh and J.~Y.~Ollitrault,
Phys. Rev. C \textbf{63}, 054906 (2001)
doi:10.1103/PhysRevC.63.054906
[arXiv:nucl-th/0007063 [nucl-th]].

\bibitem{Borghini:2001vi} 
N.~Borghini, P.~M.~Dinh and J.~Y.~Ollitrault,
Phys.\ Rev.\ C {\bf 64}, 054901 (2001)
[nucl-th/0105040].
	

\bibitem{Adler:2007aa} 
S.~S.~Adler {\it et al.} [PHENIX Collaboration],
Phys.\ Rev.\ C {\bf 77}, 014905 (2008)
[arXiv:0708.2416 [nucl-ex]].

\bibitem{Aamodt:2010cz}
K.~Aamodt \textit{et al.} [ALICE],
Phys. Rev. Lett. \textbf{106}, 032301 (2011)
doi:10.1103/PhysRevLett.106.032301
[arXiv:1012.1657 [nucl-ex]].

\bibitem{Schenke:2012wb}
B.~Schenke, P.~Tribedy and R.~Venugopalan,
Phys. Rev. Lett. \textbf{108}, 252301 (2012)
doi:10.1103/PhysRevLett.108.252301
[arXiv:1202.6646 [nucl-th]].

\bibitem{Miller:2003kd}
M.~Miller and R.~Snellings,
[arXiv:nucl-ex/0312008 [nucl-ex]].

\bibitem{Jia:2013tja}
J.~Jia and S.~Mohapatra,
Phys. Rev. C \textbf{88}, no.1, 014907 (2013)
doi:10.1103/PhysRevC.88.014907
[arXiv:1304.1471 [nucl-ex]].

\bibitem{Aad:2013xma}
G.~Aad \textit{et al.} [ATLAS],
JHEP \textbf{11}, 183 (2013)
doi:10.1007/JHEP11(2013)183
[arXiv:1305.2942 [hep-ex]].

\bibitem{Voloshin:2006gz}
S.~A.~Voloshin,
[arXiv:nucl-th/0606022 [nucl-th]].

\bibitem{Voloshin:2007pc}
S.~A.~Voloshin, A.~M.~Poskanzer, A.~Tang and G.~Wang,
Phys. Lett. B \textbf{659}, 537-541 (2008)
doi:10.1016/j.physletb.2007.11.043
[arXiv:0708.0800 [nucl-th]].


\bibitem{navid}
N.~Abbasi, D.~Allahbakhshi, A.~Davody and S.~F.~Taghavi,
Phys. Rev. C \textbf{98}, no.2, 024906 (2018)
doi:10.1103/PhysRevC.98.024906
[arXiv:1704.06295 [nucl-th]].

\bibitem{hadi}
H.~Mehrabpour and S.~F.~Taghavi,
Eur. Phys. J. C \textbf{79}, no.1, 88 (2019)
doi:10.1140/epjc/s10052-019-6549-2
[arXiv:1805.04695 [nucl-th]].


\bibitem{ALICE:2016kpq} 
J.~Adam {\it et al.} [ALICE Collaboration],
Phys.\ Rev.\ Lett.\  {\bf 117}, 182301 (2016)
[arXiv:1604.07663 [nucl-ex]].



\bibitem{Jia:2012sa} 
J.~Jia [ATLAS Collaboration],
Nucl.\ Phys.\ A {\bf 910-911}, 276 (2013)
[arXiv:1208.1427 [nucl-ex]].



\bibitem{Aad:2014fla} 
G.~Aad {\it et al.} [ATLAS Collaboration],
Phys.\ Rev.\ C {\bf 90}, no. 2, 024905 (2014)
[arXiv:1403.0489 [hep-ex]].


\bibitem{Kendall:1945}
 M.G.~Kendall,
 "The advanced theory of statistics",
 (Charles Griffinand Company, London, 1945)

\bibitem{Cramer:1999} 
H.~Cramer,
"Mathematical methods of statistics",
Princeton Mathematical Series, 
no. 9. (Princeton University Press,Prinston, 1946).

\bibitem{Krzanowski:2000} 
W.~Krzanowski,
"Principles of Multivariate Analysis",
Oxford Statistical Science Series, 
(Oxford University Press,Oxford, 2000).

\bibitem{Brenn:2017}
T.~Brenn  and  S.N.~Anfinsen,
“A revisit of the Gram-Charlier and Edgeworth series expansions,”  UiT The Arctic  University of Norway, Department of Physics and Technology, Tech. Rep., June  2017.[Online]. Available: munin

\bibitem{Giacalone:2016eyu}
G.~Giacalone, L.~Yan, J.~Noronha-Hostler and J.~Y.~Ollitrault,
Phys. Rev. C \textbf{95}, no.1, 014913 (2017)
doi:10.1103/PhysRevC.95.014913
[arXiv:1608.01823 [nucl-th]].

\bibitem{young:2009}
G.A.~Young,
M2S1 Lecture Notes,
https://wwwf.imperial.ac.uk/~ayoung/, September 2009.   

\bibitem{Mordasini:2019hut}
C.~Mordasini, A.~Bilandzic, D.~Karakoç and S.~F.~Taghavi,
Phys. Rev. C \textbf{102}, no.2, 024907 (2020)
doi:10.1103/PhysRevC.102.024907
[arXiv:1901.06968 [nucl-ex]].


\bibitem{Voloshin:1994mz}
S.~Voloshin and Y.~Zhang,
Z. Phys. C \textbf{70}, 665-672 (1996)
doi:10.1007/s002880050141
[arXiv:hep-ph/9407282 [hep-ph]].

\bibitem{Bozek:2009dw} 
P.~Bozek,
Phys.\ Rev.\ C {\bf 81}, 034909 (2010)
[arXiv:0911.2397 [nucl-th]].

\end{thebibliography}
\end{document}